\documentclass[
 preprint,
 amsmath,amssymb,
 aps, physrev,
]{revtex4-2}
\usepackage[utf8]{inputenc}
\usepackage{array}[=2016-10-06]

\usepackage{graphicx}
\graphicspath{{figures/}}
\usepackage{amsmath}
\usepackage{siunitx}

\usepackage{hyperref}
\hypersetup{
 pdftitle={A Kinetic Study of the Interactions of an Acoustic Vibrating Surface with Hypersonic Boundary-Layer Instabilities},
 pdfauthor={Mert Senkardesler, Irmak T. Karpuzcu, Deborah A. Levin}
}

\usepackage{setspace}
\usepackage{etoolbox}
\makeatletter
\patchcmd{\@makecaption}{\small\rmfamily}{\small\rmfamily\setstretch{1}}{}{\PackageError{driver}{REVTeX caption patch failed}{}}
\makeatother
\AtBeginEnvironment{table}{\setstretch{1}}
\AtBeginEnvironment{table*}{\setstretch{1}}
\usepackage{float}
\usepackage{booktabs}
\usepackage{subcaption}
\newcommand{\FNUM}{\mathit{FNUM}}

\begin{document}

\title{A Kinetic Study of the Interactions of an Acoustic Vibrating Surface with Hypersonic Boundary-Layer Instabilities}
\author{Mert Senkardesler}\email{merts2@illinois.edu}
\author{Irmak T. Karpuzcu}
\author{Deborah A. Levin}
\affiliation{University of Illinois, Urbana-Champaign, IL 61801, USA}

\date{October 21, 2026}
\begin{abstract}
The response of Mack mode instabilities to a simple acoustic vibrating surface (AVS) is studied using the Direct Simulation Monte Carlo (DSMC) method to capture and analyze spontaneous second-mode instability growth.  The DSMC simulations, analyzed with the Dynamic Mode Decomposition (DMD) method, model flow over a flat plate at an
 unseeded freestream Mach number of 6 and unit Reynolds number $Re_1\sim 1.18\times10^7$ m$^{-1}$. Power spectral density (PSD) analysis was found to  identify the dominant frequencies of 200 to 500~kHz, in line with linear stability theory (LST) predictions.  Closed-form kinetic-theory expressions were derived to predict the amplitude of the DSMC fluctuations in the window-averaged snapshots.
    It was observed that the DSMC-simulated flowfield naturally developed second Mack mode instabilities without any external seeding, {\em i.e.,} arising spontaneously from the kinetic fluctuations of the simulated gas and of amplitude higher than the predicted DSMC fluctuations.
Targeted interaction with these flow instabilities is demonstrated for an AVS, where forcing at the unstable frequency of 250~kHz results in amplified waves downstream, while at the stable frequency of 400~kHz AVS-induced disturbances are damped. The AVS thus excites only the waves at its forcing frequency, leaving the instability waves at other frequencies and the mean flow essentially unaltered. The ability of the kinetic simulations to capture and describe linear perturbations at Reynolds numbers corresponding to those of Mach~6 flight at altitudes of ${\sim}20$~km  suggests that DSMC provides a theoretical understanding  of controlling laminar-turbulent transition through AVS materials in hypersonic boundary layers.

\end{abstract}
\maketitle

\section{Introduction}

Controlling hypersonic laminar boundary-layer transition to turbulence has the potential to develop strategies that will mitigate thermal loads and aerodynamic drag to improve vehicle performance.  One possible route involves the use of phononic materials  to control boundary-layer transition through the use of phononic subsurfaces or metamaterials that would passively manipulate acoustic disturbances at the vehicle wall.\cite{brehmavs} While the material design is still in its infancy, recent  measurements of Navarro {\em et al}\cite{navarro2025} demonstrated that a bilayer phononic material subsurface reduced flow unsteadiness in a Mach 7.2 shock-wave/boundary-layer interaction  experiment, although the interaction with second-mode instabilities was not considered.   Zhao et al.~\cite{zhaoReviewAcousticMetasurfaces2022} present a review of planar metamaterial acoustic structures and their role in suppressing the second Mack mode with 
 little effect on the mean flow.   Clearly, understanding how the phononic material can modify the hypersonic boundary layer to alter Mack mode disturbance growth rates is an important component in the ultimate design of such materials.  In this work we use a novel particle-kinetic approach, in which the vibrating wall acts directly on the reflected particles, to study this fluid-surface interaction.

For Mach numbers greater than about four, an inviscid acoustic instability first identified by Mack  as the second mode prevails and governs the transition process in planar and axisymmetric high-speed boundary layer flows \cite{mackBoundaryLayerLinearStability}.  As is well known, the regions of the
 boundary layer where the local flow is supersonic relative to the disturbance phase velocity
 support selectively amplified acoustic waves that reflect between the wall and sonic line~\cite{fedorovTransitionStabilityHighspeed2011}.
 Although the phenomenon of Mack mode instability has been well studied, the use of kinetic methods to interpret this instability offers new insights due to its ability to inherently determine fluctuations  of the boundary layer base flow.
The purpose of this work is to propose a particle-kinetic treatment of  hypersonic boundary layer instabilities introduced by an externally forced source.
Instabilities are first investigated without any external seeding and  then extended to an externally forced source.  In the present work that forced source is a vibrating surface that can create external excitations that interact with quiescent boundary layer disturbances.
   First we show that an externally unseeded particle-kinetic approach is able to predict the acoustic disturbances created by the well-known Mack modes and then assess  the response of those modes to that vibrating surface. A particle-kinetic approach introduces additional momentum into the boundary layer by directly altering the velocity distributions of the reflecting particles, thereby capturing the effects of surface interactions at the micro-length scale.  Velocity distribution functions at the surface boundary are easily implemented in the Direct Simulation Monte Carlo (DSMC) kinetic approach, which enables a molecular-level model of gas-surface dynamics without the additional grid complications.

In the present effort, a kinetic approach to analyze hypersonic boundary layer instabilities is adopted using DSMC, a first-principles kinetic solver of the Boltzmann equation that captures a number of important effects:  stochastic fluctuations that deterministic Navier--Stokes solvers omit at small scales, velocity-slip and temperature-jump effects, shock-layer internal structure, and shock--boundary-layer coupling~\cite{Bird}.  DSMC has historically been used in rarefied, nonequilibrium regimes (including reacting chemistry)~\cite{gimelshein_nonequilibrium_2019,karpuzcu_collisional_2022}. However, due to its massively parallelizable algorithm it is well suited to present day  high performance computers that allow one to capture continuum flows.
In such regimes, DSMC, with a time step that resolves the mean collision time, has been used to study  the unsteadiness of shock-dominated laminar flows~\cite{tumuklu_modal_2019,KarpuzcuSideJet2023}.  While the vast majority of DSMC simulations are performed to model steady flows, the recent work of Refs.~\cite{tumukluPoF2,KarpuzcuSideJet2023,Karpuzcuspod} shows its importance in modeling unsteady flows dominated by shock--shock and shock--boundary-layer interactions. At the same time, use of kinetic methods offers insights into laminar flow instability that complement classic continuum-equations based approaches \cite{klothakisLinearStabilityAnalysis2022,sawant_etal_2022}.
Recent studies by Gallis et al.~\cite{gallisMolecularLevelSimulationsTurbulence2017, gallisTurbulenceEdgeContinuum2021} and McMullen et al.~\cite{mcmullen_navier2022, mcmullenThermalfluctuationEffectsSmallscale2023a} that employed  molecular-gas-dynamics methods to study turbulence of  Taylor--Green vortex flows demonstrated that kinetic methods were able to quantitatively reproduce canonical continuum turbulence, matching direct numerical simulation (DNS) energy spectra and dissipation rates at inertial scales~\cite{gallisMolecularLevelSimulationsTurbulence2017}.

 This paper has  two general  objectives.
First we quantify the sampled DSMC fluctuations, showing for the first time that the kinetic-fluctuation floor, which is dictated by the local thermodynamic state, number of simulated particles, and sampling parameters, may be determined  in the hypersonic boundary layer.   Any excess of the sampled DSMC fluctuations above this kinetic-fluctuation floor is then the signature of a flow instability.  Since the second-mode instabilities are identified separately from that kinetic-fluctuation floor,  the excess consists of spontaneously arising second Mack-mode waves at  Reynolds numbers and streamwise locations predicted by linear stability theory (LST).  Secondly, we
show that an acoustic vibrating surface (AVS) can interact selectively with these second-mode waves, and that a comparatively small equivalent wall deflection is sufficient to excite them above the kinetic-fluctuation floor.
We resolve second-mode instabilities in a Mach~6 boundary layer flow over a flat
plate of length $L=120$~mm, $Re_L=1.4\times10^{6}$,  up to two orders of magnitude higher Reynolds number than that in Klothakis et al. \cite{klothakisLinearStabilityAnalysis2022}.
DSMC simulations at these conditions capture spontaneously arising instabilities and exhibit a
higher-frequency spectrum in the range of dimensionless frequency $F \approx 1.5 \times 10^{-4}$–$3.0 \times 10^{-4}$, consistent with LST predictions.
 We also show that the mean flow is essentially unchanged, and that the AVS excites instabilities only at its operating frequency.

The manuscript is organized as follows. Section~\ref{sec:dsmcnoise} presents the DSMC method and the new development of the  kinetic-fluctuation floor limit of the simulated gas.
 Specifically, we discuss the convergence of the  boundary-layer simulations  considered with respect to particle number, quantify the sampled DSMC fluctuations of the flow macroparameters, and derive the kinetic-fluctuation floor against which these sampled DSMC fluctuations are compared.
 Section~\ref{sec:flowInstabilities} presents the unseeded kinetic findings  self-developing second-mode instabilities that constitute this excess in the present kinetic simulations, giving their spectral and spatio-temporal characteristics and comparing them against linear stability theory. The analysis of the stochastic, unsteady time-resolved flowfields is carried out using local Power Spectral Density (PSD) analysis for temporal characterization and global Dynamic Mode Decomposition (DMD) for spatio-temporal mode extraction \cite{Schmid2010_DMD}. After establishing the kinetic-fluctuation floor and characterizing the second-mode instabilities in the Baseline case, Section~\ref{sec:PM_model} investigates the targeted influence of
  the AVS on second-mode instabilities, assessing frequency-selective interactions with flow instabilities and comparing the amplification of the AVS-driven wave with the LST $e^N$. Conclusions are offered in Section~\ref{sec:conclusions}.  Many of the technical details developed in this research may be found in the Supplemental Material~\cite{SM}, referenced throughout the paper.

\newpage

\providecommand{\FNUM}{\mathit{FNUM}}
\providecommand{\toprule}{\hline}
\providecommand{\midrule}{\hline}
\providecommand{\bottomrule}{\hline}
\section{The DSMC Approach and Its Numerical Errors \label{sec:dsmcnoise}}

Direct Simulation Monte Carlo (DSMC) is a well-known, mature particle-kinetic stochastic method that provides reliable intrinsically unsteady numerical solutions of the Boltzmann equation of transport \cite{Bird} when appropriate cell size, time step and number of particles are employed. The Boltzmann equation is the general equation that describes the fluid motion allowing one to capture all nonequilibrium and rarefaction effects with high physical fidelity especially in flows where shocks are present.
In DSMC, gas flows are simulated by tracking the motion and collisions of numerous representative particles, with macroscopic flowfield properties obtained by averaging the microscopic properties of particles within each computational cell.
 Table~\ref{tab:params} shows the freestream conditions and DSMC numerical parameters used in this work to model the Mach 6 molecular nitrogen boundary layer flow over a flat plate.
  In Supplemental Material Sec.~\ref{sec:numconv}~\cite{SM} the convergence study performed to test the adequacy of these numerical parameters, denoted as ``Baseline'' throughout this work,  is discussed.  The DSMC numerical parameters satisfy the typical spatial resolution for collision cells as well as time step.  The time step is one third of the mean collision time (given as \(\Delta t/\tau_c\) in Table~\ref{tab:params}), and collisions are computed in adaptively refined sub-cells of the sampling grid, whose number $N_{\mathrm{coll}}$ and particle content are listed in Table~\ref{tab:params}.

To predict boundary layer instabilities, we need to calculate  fluctuations in the boundary layer once the mean flow has reached steady state.  The DSMC time step, however,   is chosen so that local collision and transport time scales are resolved, {\em i.e.,}  for boundary layer conditions, in the nanosecond range.  Boundary-layer disturbances, however,  typically  evolve over much longer periods, of order microseconds for the hundreds-of-kilohertz second-mode band.  Therefore, after the flow has reached steady state, {\em e.g.,} the 700,000 time steps (see
Table~\ref{tab:params}), we typically average macroparameters over \(N_s\) DSMC time steps for each cell of the sampling grid,
a grid which is coarser than the DSMC collision grid but sufficiently spatially resolved to capture flow
disturbances. This process, which generates a ``window snapshot,'' allows us to reduce our DSMC macroparameter
variances without the loss of information related to boundary layer fluctuations. These \(N_t\) window snapshots are then
used in the RMS and PSD analyses presented in this paper to characterize the amplitudes and frequencies of the boundary
layer disturbances. The DMD analyses, which extract the spatial modes, use a shorter time window given in Supplemental Material Sec.~\ref{sec:dmdmethod}~\cite{SM}.

\begin{table}[H]
\centering
\caption{Freestream values and DSMC numerical parameters for the Baseline case.}
\label{tab:params}
\scriptsize
\setlength{\tabcolsep}{3pt}
\begin{tabular}{@{}ll@{}p{1em}@{}ll@{}}
\hline\hline
\multicolumn{2}{@{}l}{Freestream conditions} & &
\multicolumn{2}{l@{}}{DSMC numerical and sampling parameters}\\
\cline{1-2}\cline{4-5}
Working gas & Nitrogen & & Domain & \(x=0\)--120 mm, \(y=0\)--3 mm\\
Mach number \(M_\infty\) & 6 & & Sampling grid & \(2402\times 60\)\\
Velocity \(U_\infty\) & 860.6 m/s & & DSMC time step \(\Delta t\) & 1.0 ns\\
Temperature \(T_\infty\) & 50 K & & Snapshot period \(N_s\) & 50 steps = 50 ns\\
Pressure \(p_\infty\) & 828.3 Pa & & Analysis window (time steps) & 700,000--800,000\\
Density \(\rho_\infty\) & \(5.58\times 10^{-2}\) kg/m\(^3\) & & Window snapshots \(N_t\) & 2001\\
Number density \(N_\infty\) & \(1.2\times 10^{24}\) m\(^{-3}\) & & Particle weight \(\FNUM\) & \(8.525\times 10^{10}\)\\
Mean free path \(\lambda_\infty\) & \(6.148\times 10^{-7}\) m & & Simulation particles \(N_p\) & $4.1\times 10^9$\\
Mean collision time \(\tau_c\) & 3.05 ns & & Particles per \(\lambda_\infty^2\) & 5.32\\
\(\Delta t/\tau_c\) & 0.33 & & Collision cells $N_\mathrm{coll}$ & $\approx 4.7\times 10^{8}$\\
$Re_\infty$ & \(1.176\times 10^7\) m\(^{-1}\) & & Plate length $L$ & 120 mm\\
Wall temperature \(T_w\) & 300 K & & Production resources & 80 nodes, 3840 MPI ranks\\
\hline\hline
\end{tabular}
\end{table}

Each window snapshot holds the DSMC macroparameters at each sampling grid location averaged over
\(N_s=50\) time steps. The mean-subtracted fluctuation \(q'_t\) at a time snapshot indicated by time index \(t\) and the mean sampled macroparameter \(\bar q\) over the \(N_t=2001\) window snapshots may be expressed as,
\begin{equation}
q'_t=q_t-\bar q,
\qquad
\bar q=\frac{1}{N_t}\sum_{t=1}^{N_t}q_t,
\label{eq:fluctuation}
\end{equation}
and \(q\) denotes the sampled DSMC variable of interest, such as pressure \(p\), density \(\rho\), streamwise velocity \(u\),
wall-normal velocity \(v\), temperature \(T\), or any other unsteady field quantity. Again, \(q_t\) denotes the sampled
value of that same variable at a stored time index \(t\), where the subscript \(t=1,\ldots,N_t\) labels the window snapshots
here analyzed from 700,000--800,000 DSMC timesteps. With this definition, the DSMC RMS
fluctuations are therefore,
\begin{equation}
q_{\mathrm{rms}}=
\sqrt{\frac{1}{N_t}\sum_{t=1}^{N_t}\left(q'_t\right)^2}
=
\sqrt{\frac{1}{N_t}\sum_{t=1}^{N_t}\left(q_t-\bar q\right)^2}.
\label{eq:rms}
\end{equation}

Figure~\ref{fig:rmsfields} of the Supplemental Material~\cite{SM} shows the spatial distribution of the steady state fluctuation values for the macroparameters of density, temperature, pressure, and streamwise velocity.   At this point it is not possible to separate those fluctuations due to numerical noise from those that are physical boundary layer instabilities.  While the effect of DSMC sample size follows typical Poisson statistics, correlation effects that can occur in the  time averaging used to form window snapshots must also be considered.    Therefore, the degree of correlation must be quantified to obtain accurate predictions of the {\it numerical} errors in the DSMC RMS values in hypersonic boundary layer flows.

Prior work has partially addressed this.
Hadjiconstantinou et al.~\cite{Hadjiconstantinou2003} derived equilibrium sampling variances for particle estimates of hydrodynamic observables,
providing the one-snapshot coefficients for count, velocity, temperature, and pressure used in Eqs.~\eqref{eq:rho_theory}--\eqref{eq:p_theory}. Garcia~\cite{Garcia2006} showed
that hydrodynamic variables must be formed from mechanical densities because these observables are ratios or nonlinear
functions of finite particle sums. Temporal correlation between successive DSMC samples has also been analyzed
through persistent-random-walk, Markov-process, modified central-limit-theorem, autocorrelation, and flux-based
exchange models~\cite{Hadjiconstantinou2003,PlotnikovShkarupa2014,PlotnikovShkarupa2015,PlotnikovShkarupa2016,GalitzineBoyd2015,Li2016}. In most of this literature, however, the target quantity is the convergence of a long-time
sampled mean, either through the decay of its statistical error or through the sample count required to achieve a
prescribed confidence interval.   Galitzine and Boyd~\cite{GalitzineBoyd2015} obtained  finite
correlated-sample covariance identities, but their  expressions corresponded to the large-window
integrated-correlation limit and their general framework required a measured autocorrelation spectrum.
  Using these earlier contributions we show that
 when both the finite number of computational particles and the correlations due to sampling in
the hypersonic boundary layer are considered, the predicted fluctuations are given as,
\begin{align}
\frac{n'_{\mathrm{theory,rms}}}{\bar n}
&=
\frac{\rho'_{\mathrm{theory,rms}}}{\bar \rho}
=\epsilon_n,
\label{eq:rho_theory}\\
u'_{\mathrm{theory,rms}}
&=
\sqrt{\frac{k_B\bar T}{m}}\,\epsilon_n,
\label{eq:u_theory}\\
\frac{T'_{\mathrm{theory,rms}}}{\bar T}
&=
\sqrt{\frac{2}{f}}\,\epsilon_n,
\label{eq:t_theory}\\
\frac{p'_{\mathrm{theory,rms}}}{\bar p}
&\approx
\sqrt{1+\frac{2}{f}}\,\epsilon_n,
\label{eq:p_theory}
\end{align}
where
\begin{equation}
\epsilon_n:=
\frac{1}{\sqrt{\bar N_{p,\mathrm{cell}}\alpha_n N_s}}.
\label{eq:epsilon}
\end{equation}
The finite-window independence ratio in Eq.~\eqref{eq:epsilon} is evaluated from the local particle-renewal
estimate as
\begin{equation}
\resizebox{0.98\linewidth}{!}{$\displaystyle
\alpha_n(N_s;\tau,\bar{\mathbf U},a)
=
\left[
1+2\sum_{\ell=1}^{N_s-1}
\left(1-\frac{\ell}{N_s}\right)
\left(
1-\frac{\Delta t}{\Delta x\Delta y\Delta z}
\left[
\begin{aligned}
&\big(G_+(\bar u_x,a)+G_-(\bar u_x,a)\big)\Delta y\Delta z\\
&+\big(G_+(\bar u_y,a)+G_-(\bar u_y,a)\big)\Delta x\Delta z
\end{aligned}
\right]
\right)^{\ell}
\right]^{-1},
$}
\label{eq:alpha_expanded}
\end{equation}
where $\Delta t$ is the DSMC timestep, \(\Delta x\) and \(\Delta y\) are the sampled-cell dimensions in the resolved
streamwise and wall-normal directions, and \(\Delta z\) is the unit spanwise sampling depth used to form the sampled
cell volume and face areas.
\(\bar{\mathbf u}=(\bar u_x,\bar u_y)\) is the local resolved bulk velocity, and \(a=\sqrt{k_B\bar T/m}\) is the one-dimensional
thermal speed. The functions \(G_+\) and \(G_-\) are the positive and negative half-range molecular crossing speeds
through the corresponding cell faces, computed from the local shifted Maxwellian in that coordinate direction.
Also in the above equations,
\(n\) is the number density, \(m\) the particle mass, \(k_B\) Boltzmann's constant, and \(f\) is the number of
active thermal degrees of freedom. For nitrogen with translational and rotational modes active, \(f=5\). The quantities \(\bar N_{p,\mathrm{cell}}\), \(\alpha_n\), and
\(N_s\) are the number of particles per sampling cell, per time step, the independence ratio, and the number of DSMC time
step samples averaged over (here \(N_s=50\)), respectively. The independence ratio takes on a value between 0 and
unity, with the upper limit corresponding to a case of no correlations. Since the DSMC sampling cells are the
same as those used in the window snapshots, \(\bar N_{p,\mathrm{cell}}\) is between \(10^4\)--\(10^5\) particles per
sampling cell, per DSMC time step.  For the present conditions \(\alpha_n\approx0.02\), that is, over the 50 consecutive time samples from a cell, it holds largely the same particles, unchanged, so the samples are far from independent.
Equation~\eqref{eq:epsilon} shows that the theoretical DSMC error is inversely proportional to the square root of both the sample size and the
fraction of independent samples.

These expressions unify the two factors that contribute to DSMC numerical noise, i.e.,
finite particle count and the presence of relatively slow-moving flow even in a compressible flow boundary layer leading
to correlations.
With the above treatment we will now be able to demonstrate in Secs.~\ref{sec:flowInstabilities} and \ref{sec:PM_model} that the  fluctuations predicted from the DSMC-driven PSD and DMD analyses of the Mack modes and AVS disturbances are well above the predicted DSMC kinetic fluctuations, Eqs.~\eqref{eq:rho_theory}--\eqref{eq:epsilon}, especially in the spatial and frequency regions of greatest importance. This quantified kinetic floor is akin to that of
 thermal fluctuations  in the receptivity analysis of Fedorov and Tumin~\cite{fedorovReceptivityHighSpeedBoundary2017}, set within the fluctuating hydrodynamics of Landau and Lifshitz~\cite{landauHydrodynamicFluctuations1957,ortizdezarateHydrodynamicFluctuations2006}.  However, in that analysis the fluctuations enter as stochastic forcing at the intensity fixed by the fluctuation--dissipation theorem, and the instability amplitude follows in closed form. In the present work the fluctuations are intrinsic to the simulation, at the level set by Eqs.~\eqref{eq:rho_theory}--\eqref{eq:epsilon}, and the second-mode waves arise from them without imposed forcing.

\section{Kinetically Excited Second-Mode Instabilities}
\label{sec:flowInstabilities}

In hypersonic boundary layers,
at the Mach 6 conditions of this study, the first acoustic mode, referred to as the second mode, becomes the dominant instability, exhibiting significantly larger amplification rates \cite{mackBoundaryLayerLinearStability}.
As is well known, a powerful acoustic resonance mechanism explains the dominance of the second mode in hypersonic flows and, as will be seen in the DSMC simulations, accounts for the strongly amplified frequencies in the 200--500~kHz range.

\begin{figure}[H]
    \centering
    \includegraphics[width=\linewidth,height=0.9\textheight,keepaspectratio]{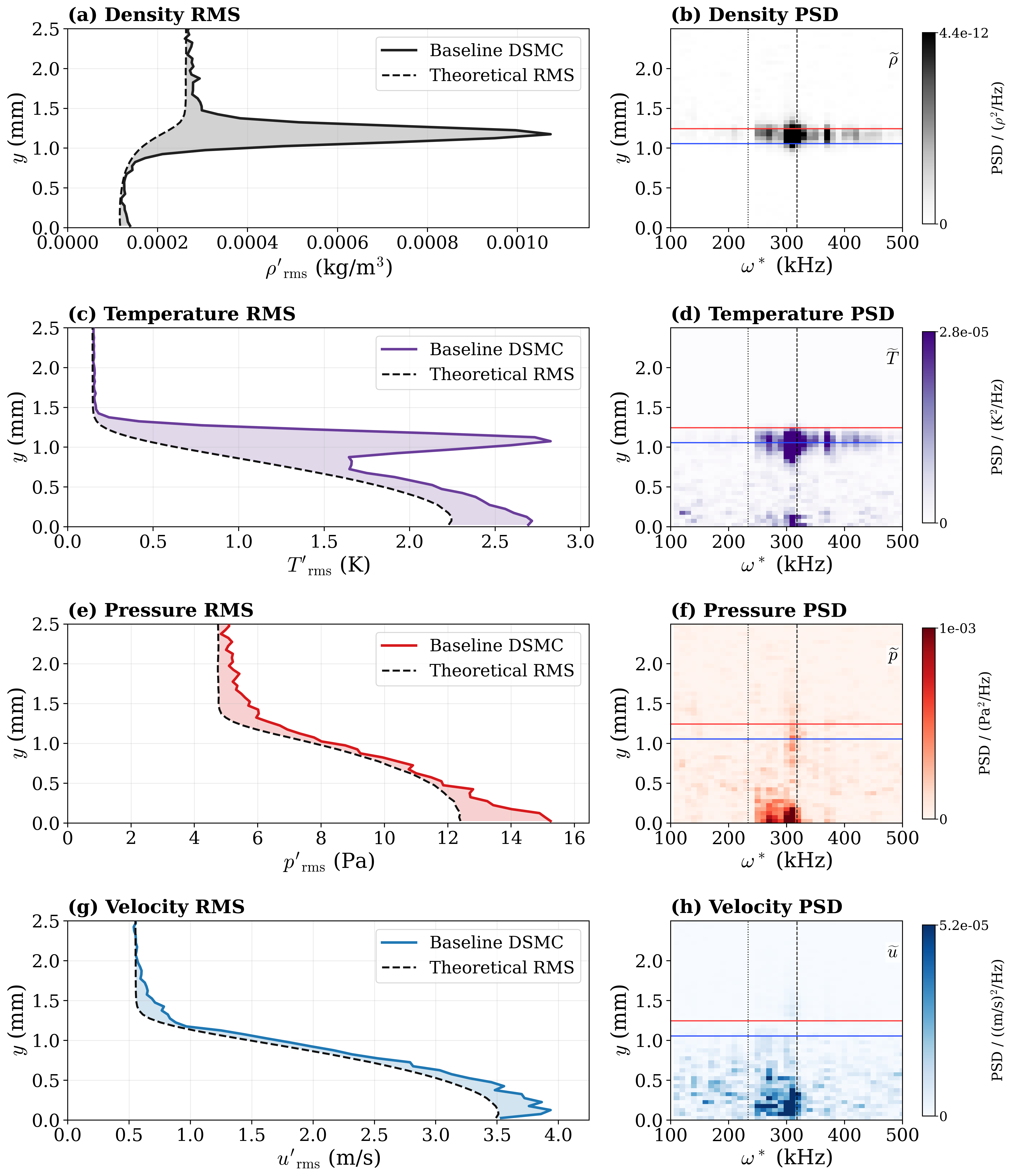}
    \caption{Aligned RMS-excess and PSD diagnostics for density, temperature, pressure, and streamwise velocity at $x=80$~mm. The left column (a,c,e,g) compares the measured DSMC RMS profiles against the theoretical kinetic-noise RMS (Eqs.~\eqref{eq:rho_theory}--\eqref{eq:epsilon}). The shaded area is the measured excess above this theoretical level. The right column (b,d,f,h) shows the corresponding wall-normal PSD profiles at the same location. The dotted and dashed vertical guides mark the frequencies of the second-mode identity location and the upper neutral point at this $x$, $\omega^*\simeq234$ and $318$~kHz, and the blue and red horizontal lines mark the inflection point and the boundary-layer height.}
    \label{fig:psd_bl_profiles}
\end{figure}

As discussed in the previous section, the predicted theoretical kinetic-fluctuation floor lets us separate the physical, amplified instabilities from the DSMC noise.  The left column of Fig.~\ref{fig:psd_bl_profiles}(a,c,e,g) along the boundary-layer height at $x=80$~mm shows the measured DSMC RMS of density, temperature, pressure, and streamwise velocity (solid lines) together with the theoretical RMS of the kinetic fluctuations of the respective variable (dashed lines, Eqs.~\eqref{eq:rho_theory}--\eqref{eq:epsilon}). The shaded area is the measured excess RMS above this theoretical level.
The accompanying PSD profiles in the right column of Fig.~\ref{fig:psd_bl_profiles}(b,d,f,h) as a function of frequency and wall-normal coordinate are computed with a Hann window over the full 100~$\mu$s record, from the $N_t=2001$ window snapshots with a sampling frequency of 20~MHz, giving a frequency bin spacing of 10~kHz. The PSD profiles place the spectral energy at the same wall-normal heights as the RMS excess. The excess RMS and the PSD peaks thus come from the same fluctuations, whose energy is concentrated in a specific frequency band, which in this case is that of the second-mode instability. The kinetic fluctuations inherent to DSMC therefore not only excite the instability but also set the reference level against which it is measured (Eqs.~\eqref{eq:rho_theory}--\eqref{eq:epsilon}).

In density and temperature [Fig.~\ref{fig:psd_bl_profiles}(a,c)], the RMS exceeds the kinetic level mainly along the shear layer near the boundary-layer edge, with a smaller near-wall excess in temperature. The PSD profiles in Fig.~\ref{fig:psd_bl_profiles}(b,d) place the strongest fluctuations at exactly these heights, around the inflection point and the boundary-layer edge (blue and red horizontal lines). Their energy lies within the locally unstable second-mode band from LST (Supplemental Material Sec.~\ref{sec:secondmacmode}~\cite{SM}), marked by the vertical black lines, and peaks near its upper limit, consistent with unstable frequencies that originate upstream and convect to the measurement location. In pressure and streamwise velocity [Fig.~\ref{fig:psd_bl_profiles}(e,g)], the excess lies mainly toward the wall, and the PSD profiles in Fig.~\ref{fig:psd_bl_profiles}(f,h) show the fluctuations at the same near-wall heights, again within the unstable band.

\begin{figure}[H]
    \centering
    \includegraphics[width=\linewidth]{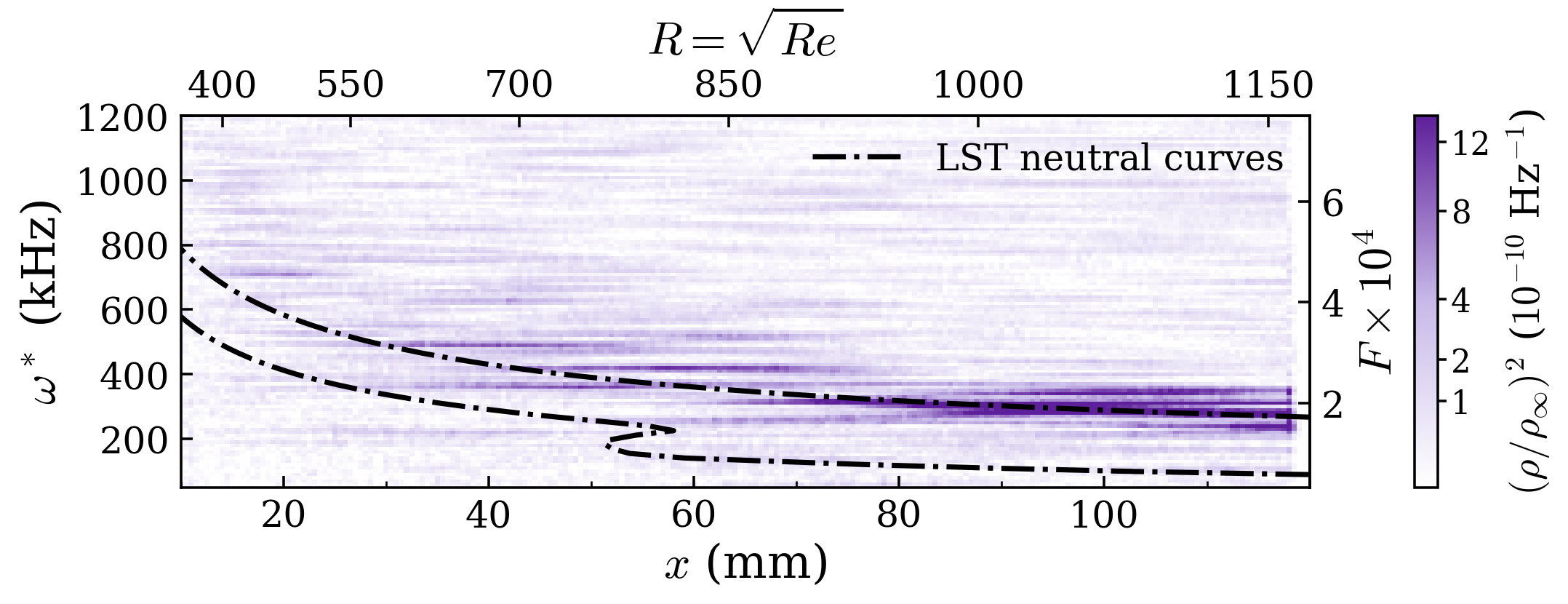}
    \caption{Spatial distribution of the density spectra along the boundary-layer edge, $y=\delta_{99}(x)$, over $x=10$--$120$~mm, with the LST second-mode neutral curves overlaid.}
    \label{fig:edge_psd}
\end{figure}

The present 120~mm long domain is short, corresponding to a Reynolds number $R=\sqrt{Re}\approx 1170$ at its end (Supplemental Material Eq.~\eqref{eq:ReynoldsAndLength}~\cite{SM}). Therefore, it captures only the early stages of linear growth. However, it does span a broad range of unstable frequencies, from about 200 to 500~kHz. Figure~\ref{fig:edge_psd} compares the spectra obtained from the DSMC with linear stability theory, mapping the density PSD along the boundary-layer edge, probed at each streamwise location at the local height $\delta_{99}$, with the LST neutral curves overlaid. The top and right axes give the Reynolds number $R$ and the dimensionless frequency $F$ in the convention of Mack~\cite{mackBoundaryLayerLinearStability} (Supplemental Material Sec.~\ref{sec:secondmacmode}~\cite{SM}). It can be seen that the kinetic fluctuations excite the whole band of unstable frequencies, and the spectral energy inside the unstable region grows downstream, as linear stability theory predicts. At each location the most energetic frequency lies close to the upper neutral curve.

\begin{figure}[H]
    \centering
    \includegraphics[width=\linewidth]{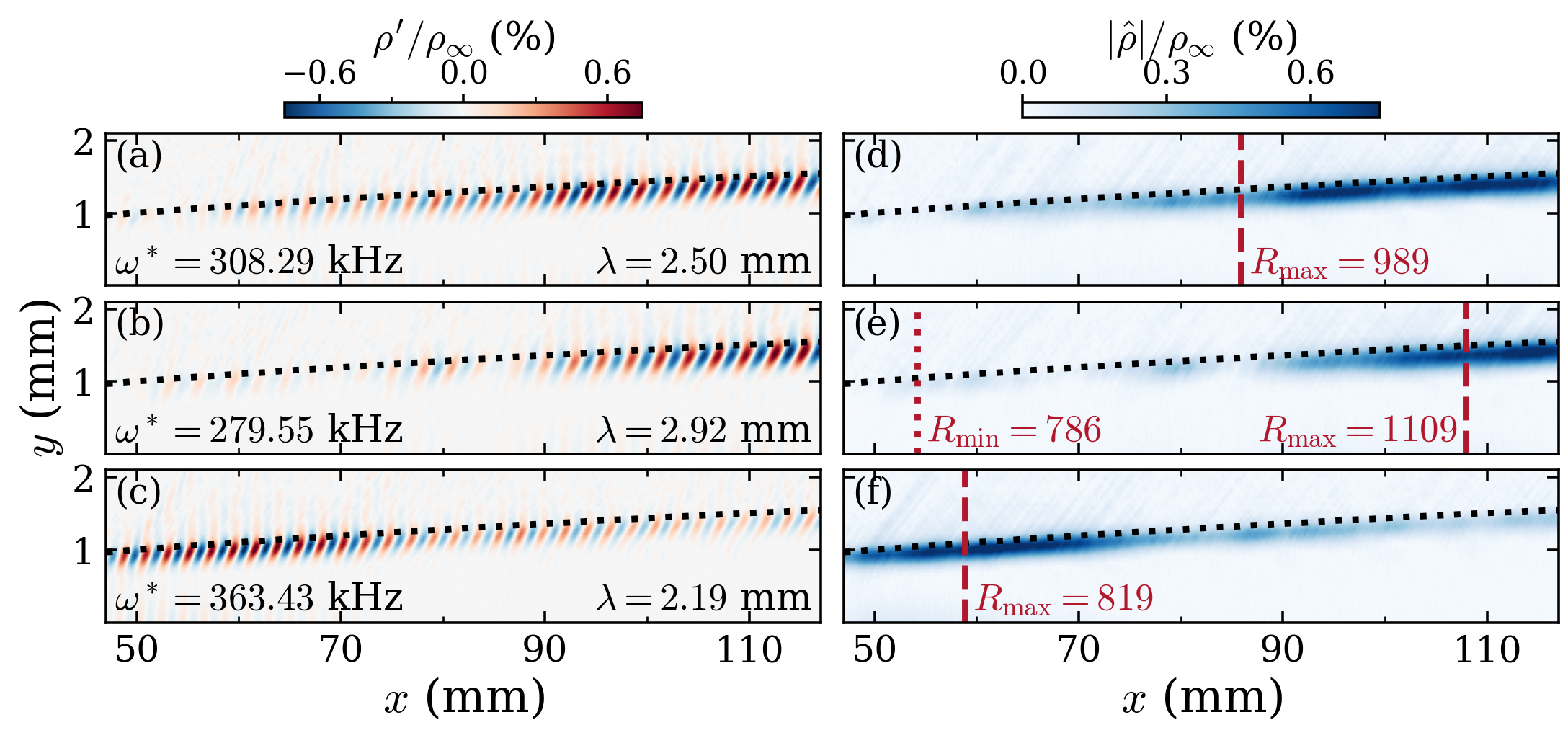}
    \caption{DMD spatial distribution of the three most energetic density modes of the Baseline field, capturing the propagating second-mode wave structures: real parts of the DMD modes (a--c) and modal amplitudes $|\hat{\rho}|/\rho_\infty$ (d--f), where $|\hat{\rho}|$ is the modulus of the complex DMD mode. (a, d) $308$~kHz, (b, e) $280$~kHz, and (c, f) $363$~kHz, with corresponding wavelengths $\lambda = 2.50$, $2.92$, and $2.19$~mm. Perturbations are given as percentages of the freestream density. The black dotted line marks the boundary-layer edge. In (d--f), the red dotted vertical line marks the second-mode identity location $R_{\min}$ and the red dashed line the upper neutral point $R_{\max}$ of each wave.}
    \label{fig:all_modes}
\end{figure}

While the local PSD analysis  effectively identifies the frequencies of second-mode disturbances at discrete points, it does not leverage the fact that these instabilities inherently exhibit spatial coherence due to their wave-like propagation within the flow. Since these locally unstable frequencies are convected and accumulated as they propagate downstream, one should expect them to manifest as spatially coherent structures, particularly in the streamwise direction.  Therefore, we next use the DMD coherent spatial modes \cite{Schmid2010_DMD,Karpuzcuspod,Ghannadian2024,senkardeslerDSMCStudyHypersonic2026} that emerge from time-series snapshots of DSMC data in the boundary layer (Supplemental Material Sec.~\ref{sec:dmdmethod}~\cite{SM}).
Figure~\ref{fig:all_modes} shows the spatial distribution of the first three modes with the greatest amplitudes predicted in the DMD analysis, corresponding to frequencies of 308, 280, and 363~kHz, in order of decreasing amplitude (see hatching in Fig.~\ref{fig:DMD_modes_freq} of the Supplemental Material~\cite{SM}).  Figure~\ref{fig:all_modes}(a--c) shows that each mode displays coherent wave patterns with well-defined wavelengths.    As shown in Fig.~\ref{fig:all_modes}, the DMD analysis successfully isolates the spatially coherent wave structures of the most energetic modes, each associated with a distinct frequency. The three density modes are concentrated near the boundary-layer edge (dotted line), with wavelengths from  2.19 to 2.92~mm for the different frequency components of the second-mode waves. Their wavelengths are approximately twice the local boundary-layer thickness, consistent with the rough estimate for second-mode waves.

The amplitudes of the DMD modes are depicted in Fig.~\ref{fig:all_modes}(d--f). For each wave, the vertical lines mark two locations from linear stability theory (Supplemental Material Sec.~\ref{sec:secondmacmode}~\cite{SM}): $R_{\min}$, the second-mode identity location, and $R_{\max}$, the upper neutral point, downstream of which LST predicts decay. Each mode grows between these two locations, as LST predicts. The 363 and 280~kHz modes peak around their upper neutral points, within a few millimeters of $R_{\max}$, and the captured 308~kHz wave is in reasonable agreement as well, reaching its maximum farther downstream, near $x\approx110$~mm.

\begin{figure}[H]
\centering
\includegraphics[width=\linewidth]{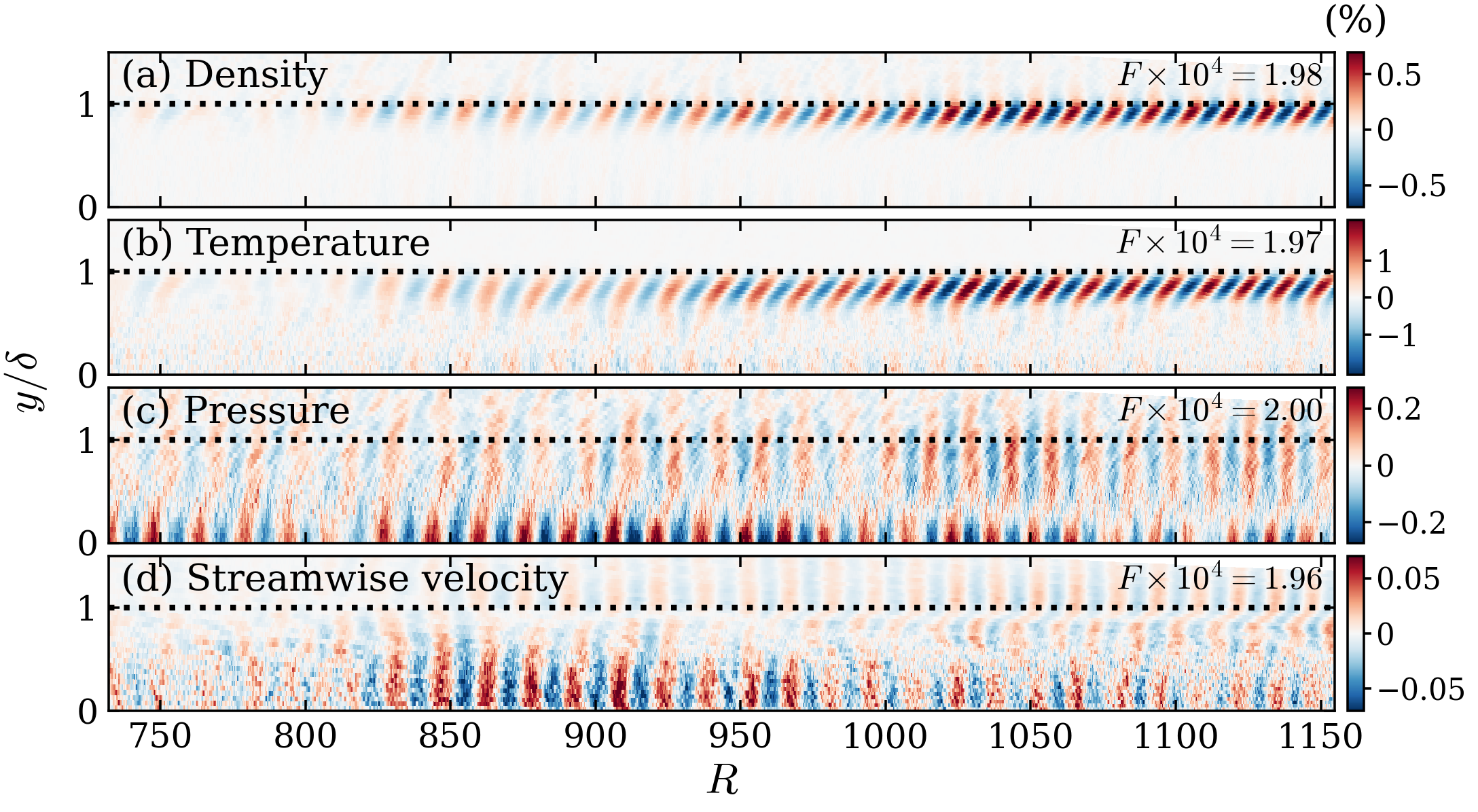}
\caption{DMD modes (real parts) for (a) density, (b) temperature, (c) pressure, and (d) streamwise velocity, highlighting coherent structures near 308~kHz. Each map is in percent of the freestream value of its variable. Panels (a--c) show mode 1 of each decomposition, and panel (d) shows mode 4 of the streamwise-velocity decomposition, at 305~kHz.}
\label{fig:mode1_only}
\end{figure}

Figure~\ref{fig:mode1_only} shows the DMD modes of density ($\rho_t$), temperature ($T_t$), pressure ($p_t$), and streamwise velocity ($u_{x,t}$) for the second-mode wave near 308~kHz, $F\approx2\times10^{-4}$. To facilitate direct comparison with linear stability predictions, DMD analysis results are converted into dimensionless coordinates, with
the wall-normal coordinate normalized by the local boundary layer thickness, $y/\delta$.
The spatial structures of these modes reveal distinct behaviors for each flow variable. For example, the density perturbation, \(\rho_t\) [Fig.~\ref{fig:mode1_only}(a)], is primarily concentrated within the shear layer near the boundary-layer edge, which is consistent with the RMS and PSD profiles in Fig.~\ref{fig:psd_bl_profiles}(a,b).
These spatially coherent wave structures captured in the density perturbations are the numerical analogue of the characteristic ``rope-like" structures frequently observed in experimental flow visualizations of second-mode instabilities using techniques such as schlieren imaging~\cite{Miller2022} and nanoparticle-tracer-based planar laser scattering (NPLS)~\cite{Zhang2023}. 

In contrast, the pressure fluctuation, \({p_t}\) [Fig.~\ref{fig:mode1_only}(c)], is most prominent near the wall, again aligning with the local RMS and PSD profiles [Fig.~\ref{fig:psd_bl_profiles}(e,f)]. Across the wall-normal direction, at any streamwise location, the pressure waves undergo a phase reversal at about \(y\approx0.4\,\delta\), which is itself a signature of second-mode acoustic waves~\cite{mackBoundaryLayerLinearStability}. Finally, the streamwise velocity perturbation, \({u_{x,t}}\) [Fig.~\ref{fig:mode1_only}(d)], is concentrated in the lower half of the layer, near the wall like the pressure perturbation. The same second-mode wave, near 308~kHz, thus shows up differently in each flow variable. Its pressure and streamwise-velocity components are established and amplified near the wall within the LST unstable region. Its density and temperature components reflect the accumulation of these locally amplified disturbances as they convect downstream, and appear most strongly in the shear layer farther downstream.

\section{Excitation of Second Mack Mode Instabilities by an Acoustic Vibrating Surface  \label{sec:PM_model}}

Having established the ability of DMD using time series data from the kinetic DSMC formulation to model second Mack mode instabilities in the 200 to 500~kHz region, we now examine how an oscillating acoustic surface placed on a flat plate will interact with these disturbances.  A natural question emerges as to whether these instabilities can be selectively excited in a predictable manner consistent with LST, with the forced response measured against the kinetic-fluctuation floor and the no-AVS reference.
 We consider the harmonic vibrational motion of an AVS as the potential source of interaction with the disturbances discussed in the previous sections, {\em i.e.,} the  AVS can be described by its time-dependent position $\widetilde{X}_s(t)$ and velocity $\widetilde{V}_s(t)$:
\begin{equation}
    \widetilde{X}_s(t) = |X_s| \sin(\omega t), \quad \widetilde{V}_s(t) = |V_s| \cos(\omega t)
\end{equation}
where \( |X_s| \) is the equivalent deflection amplitude, $|V_s| = \omega |X_s|$ is the surface velocity amplitude, and \( \omega \) is the angular frequency.

 At these high frequencies, the face velocity of the oscillating AVS influences the flow by altering the velocity of particles due to wall collisions. At a frequency of 400~kHz and a face-velocity amplitude of 50~m/s, the corresponding equivalent surface deflection is $|\widetilde{X}_s| \approx 19.9~\mu$m, about three times the local mean free path near the wall ($\lambda_w \approx 6~\mu$m, roughly an order of magnitude larger than the freestream value of $0.6~\mu$m), and 1.8 and 2.9 percent of the local boundary-layer thickness ($\delta_{99} \approx 1.08$~mm at the AVS) for the 19.9 and 31.8~$\mu$m deflections of the two 50~m/s cases. This face velocity, in turn, produces particles with new velocities in the tail of thermal distributions at the wall temperature of 300~K.
The model therefore represents the AVS only through its effect on the particles reflected from it: the interaction of the AVS with the flow is modeled by altering the velocity distribution of these reflected particles.

The AVS is therefore modeled as a prescribed normal-velocity shift applied to molecules reflected from a fixed surface. The model builds on diffuse reflection from stationary and moving walls~\cite{Bird,sone2007molecular} and on DSMC moving-boundary algorithms~\cite{raderDSMCMovingBoundary2011}, but the surface is never displaced.  An analytic expression for the distribution of normal velocity components of gas particles reflecting from the AVS surface is used to determine the
 post-collisional normal velocities  for a stationary wall, \( F(V_n^*) \),
\begin{equation}
	F(V_{n}^*) = \left( \frac{m}{k_B T_w} \right) V_{n}^* \exp\left( -\frac{m(V_{n}^*)^2}{2k_B T_w} \right), \quad 0 < V_n^* < \infty
	\label{eq:myEq1}
\end{equation}
where \( m \) is the particle mass, \( k_B \) is Boltzmann's constant, \( T_w \) is the wall temperature, and \( V_{n}^* \) is the normal velocity of the reflected particles with respect to the wall.
During the DSMC simulation, the velocity distributions of the reflected particles are shifted by the AVS's surface velocity at the time of collision giving a new
probability distribution of normal velocities \( V_n' \) for particles reflecting from a moving wall with surface velocity \( V_s \) of
\begin{equation}
    F({V_n}^{'}) = F(V_n^* - V_s^*)
\end{equation}
While only the normal velocity component is directly altered by the wall interaction, this change modifies the particle speed or its kinetic energy, which is then   redistributed throughout the flow via subsequent inter-particle collisions, propagating new AVS disturbances.   Supplemental Material Sec.~\ref{sec:derivation}~\cite{SM} provides the derivation of the reflected-particle speed distribution $f(C)$ as well as a comparison of how the Maxwellian distribution shifts relative to a non-moving surface.
To model this interaction of an AVS  time-oscillating wall
in DSMC, we adjust the normal velocities of particles reflecting from the wall by the instantaneous velocity of the AVS in its harmonic motion at the moment of collision. The harmonic surface velocity \( \tilde{V}(t) \) is utilized where particles reflecting  from the surface have their reflected normal velocities altered based on the current velocity of the AVS.
\begin{equation}
	{V}_{n}^{AVS}(t) = {V}_{n}^* + |V_s| \cos\left(\frac{2\pi t}{T} \right)
	\label{eq:my_equationAVS}
\end{equation}
 The algorithm is implemented in the DSMC framework by first sampling the stationary normal velocity ${V}_{n}^*$, corresponding to a normal velocity value for a particle reflecting from a stationary wall, and then shifting it by the current face velocity of the AVS, which is $|V_s| \cos(\frac{2\pi t}{T})$, with this velocity shift depending on the current time $t$ of the DSMC simulation and the period $T$ of the AVS's harmonic motion.
The AVS's motion is modeled in the  four forced DSMC simulations as an unsteady velocity boundary condition applied at an undeflected surface position of 60~mm from the leading edge, without actually moving the wall panel. The harmonic surface velocity amplitude, \( |V_s| =  \omega  |X_s|\), is set to 50 and 100~m/s
 with a surface plate width of 1~mm and a surface wall temperature and energy accommodation coefficients of 300~K and 1.0 (fully diffuse), respectively are assumed.

\begin{table}[H]
\centering
\small
\caption{AVS cases considered in the present study. The AVS strip spans $x=59.5$--$60.5$~mm, and the corresponding equivalent deflection amplitude is $|X_s| = |V_s|/(2\pi\,\omega^*)$.}
\label{tab:avs_cases}
\setlength{\tabcolsep}{4pt}
\renewcommand{\arraystretch}{1.1}
\begin{tabular}{lccl}
\hline\hline
Case & $\omega^*$ (kHz) & $|V_s|$ (m/s) & $|X_s|$ ($\mu$m)  \\
\hline
No AVS reference & none & 0 & 0  \\
250~kHz, 50~m/s & 250 & 50 & 31.83  \\
250~kHz, 100~m/s & 250 & 100 & 63.66  \\
400~kHz, 50~m/s & 400 & 50 & 19.89  \\
400~kHz, 100~m/s & 400 & 100 & 39.79  \\
\hline\hline
\end{tabular}
\end{table}

\begin{figure}[H]
\centering
\includegraphics[width=0.68\linewidth]{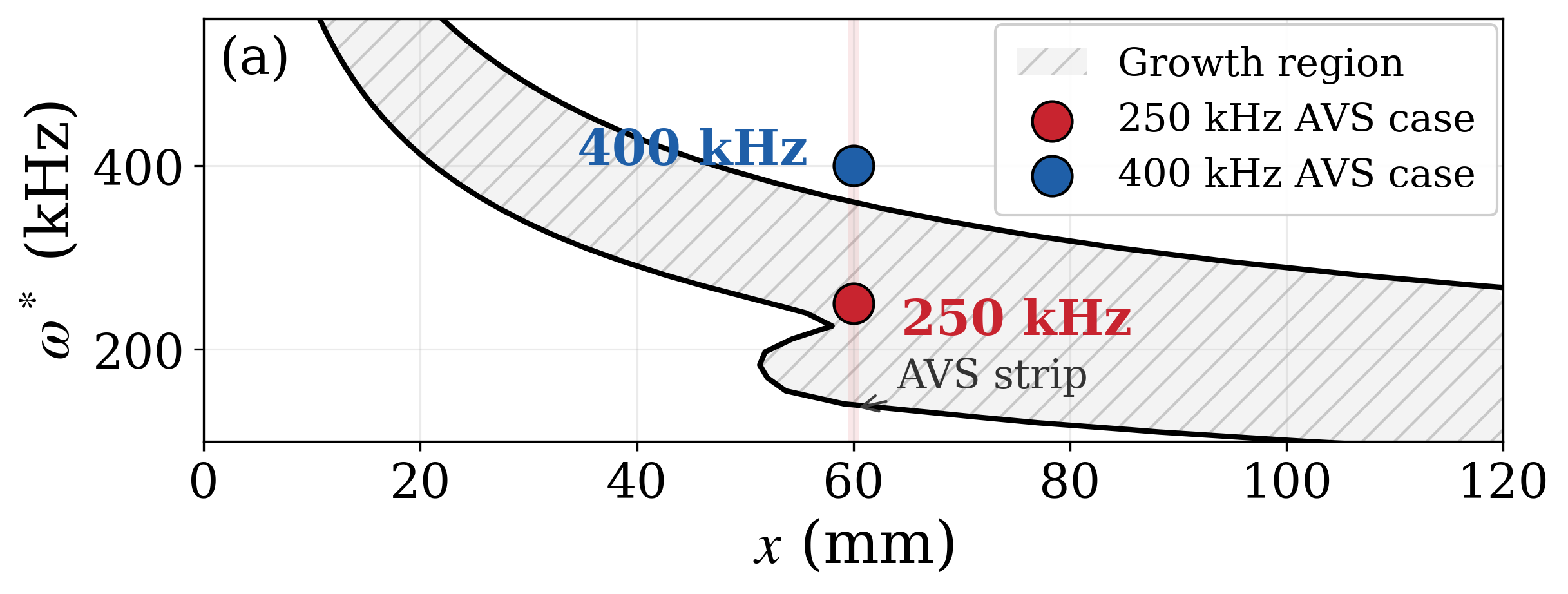}\\[-2pt]
\includegraphics[width=\linewidth]{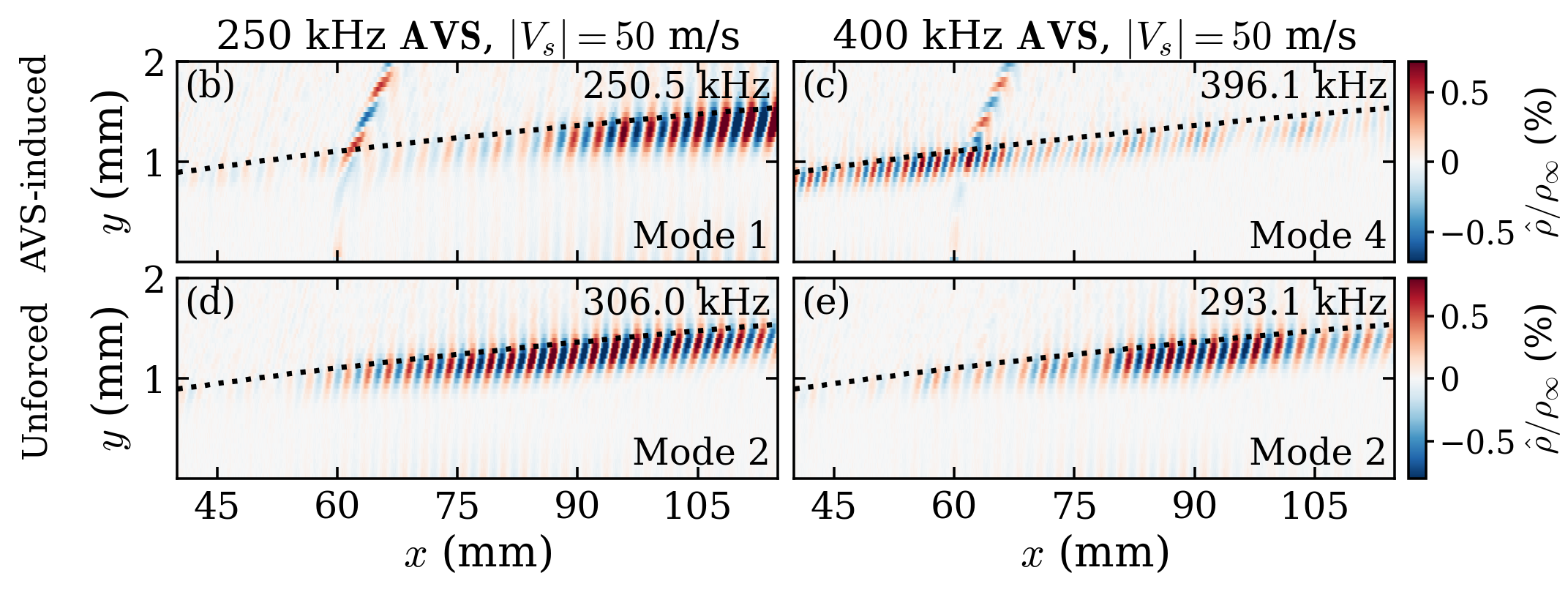}
\caption{(a) Imposed AVS frequencies on the LST second-mode neutral curves for the flat-plate mean-flow reference. (b--e) Density DMD modes for the two $|V_s|=50$~m/s AVS cases: (b, c) the AVS-induced modes nearest the forcing frequency and (d, e) the unforced second-mode waves, which arise from the kinetic fluctuations of the simulated gas. The dotted curve marks the boundary-layer edge.}
\label{fig:avs_density_modes}
\end{figure}

  The full set of cases is summarized in Table~\ref{tab:avs_cases}, in which each frequency is driven at surface-velocity amplitudes of $50$ and $100$~m/s.  For the two 50~m/s cases, frequencies of 250 and 400~kHz, the equivalent face displacement,     $\lvert X_s \rvert$
 has values of 31.8 and 19.9~$\mu $m, respectively.   These two driving frequencies
 are chosen deliberately with respect to the LST neutral curve at the forcing location $x = 60$~mm.  As Fig.~\ref{fig:avs_density_modes}(a) shows, 
the 250~kHz case lies on the lower branch, inside the unstable region, so that the AVS-seeded disturbance is expected to grow monotonically downstream, whereas the 400~kHz case lies marginally outside the unstable region and is therefore expected to decay immediately. The $50$~m/s cases will be shown to be more than sufficient to excite and interact with the second mode. The $100$~m/s cases are designed to test the selectivity of the AVS even at a higher amplitude; {\em i.e.,} to test whether the disturbances it induces at the stable frequency are still quenched, and to verify that its effect on the mean flow remains negligible for both frequencies (see Supplemental Material Sec.~\ref{sec:avsmeanflow}~\cite{SM}).

DMD is next applied to the unsteady DSMC fields for the two $50$~m/s AVS cases. The forcing starts at timestep 600,000 and the DMD window of 1001 snapshots uses the DSMC data from 750,000 to 800,000 timesteps, {\em i.e.,} after the onset of forcing.   Figure~\ref{fig:avs_density_modes}(b, c) shows the real part of the density mode closest to the forcing frequency. The $250$~kHz forcing, which lies inside the second-mode unstable band (see Fig.~\ref{fig:avs_density_modes}(a)), creates a coherent wave that grows along the boundary-layer edge downstream of the strip, confirming that the AVS excites the second mode.  In contrast,  the $400$~kHz mode, which lies outside the band at and downstream of the strip, decays immediately downstream of it, consistent with LST~\cite{mackBoundaryLayerLinearStability}.  Alongside the driven $250$~kHz mode, the DMD recovers a coherent second-mode wave near $306$~kHz [Fig.~\ref{fig:avs_density_modes}(d)], and alongside the driven $400$~kHz mode a wave near $293$~kHz [Fig.~\ref{fig:avs_density_modes}(e)]. Both are unforced second-mode waves, which arise from the    kinetic fluctuations of the simulated gas. This is consistent with the unforced baseline case [Fig.~\ref{fig:all_modes}(a)], in which these fluctuations are the only source of receptivity.
The AVS therefore strongly excites the waves at its forcing frequency, while having minimal effect on the unstable waves at the other frequencies, which persist and remain relatively unaltered.

\begin{figure}[H]
\centering
\includegraphics[width=\linewidth]{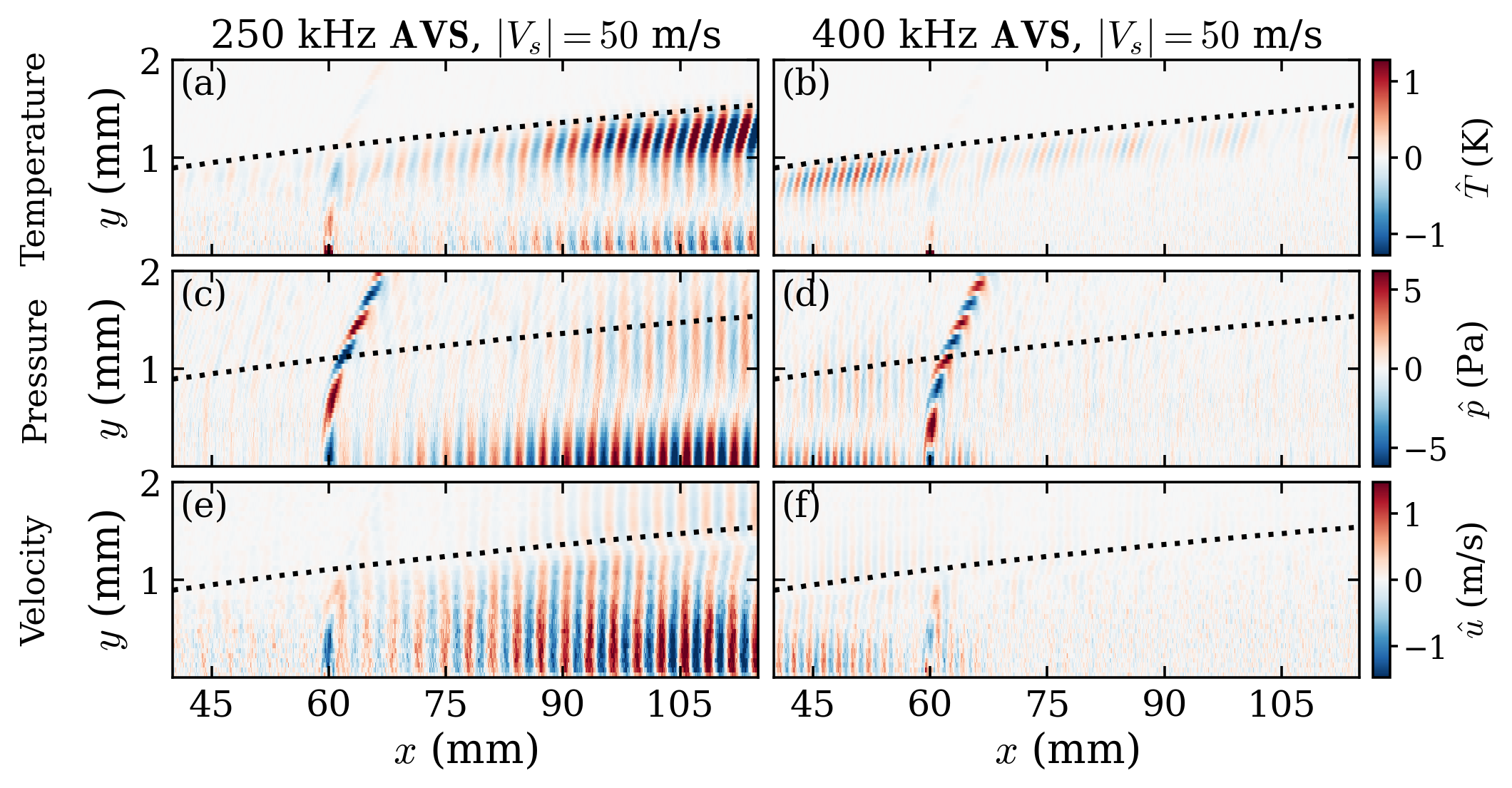}
\caption{Real part of the DMD mode closest to the imposed AVS frequency for the two $|V_s|=50$~m/s cases, in (a, b) temperature, (c, d)~pressure, and (e, f)~streamwise velocity, with the 250~kHz case on the left.  The dotted curve marks the boundary-layer edge.
}
\label{fig:avs_driven_modes_puT}
\end{figure}

The same selective response is recovered in the temperature, pressure, and streamwise-velocity fields.  In each variable, the DMD mode nearest the imposed frequency reproduces the coherent $250$~kHz wave that grows downstream, while the $400$~kHz response carries upstream content, peaks locally at the strip, and decays downstream (see Fig.~\ref{fig:avs_driven_modes_puT}). As expected for the second mode, each variable establishes its disturbance at a different wall-normal location with the pressure fluctuation being  largest near the wall, the streamwise velocity  spread over the inner boundary layer and weak at the edge, and the temperature peaking toward the boundary-layer edge with a secondary maximum near the wall.

\begin{figure}[H]
\centering
    \includegraphics[width=\linewidth]{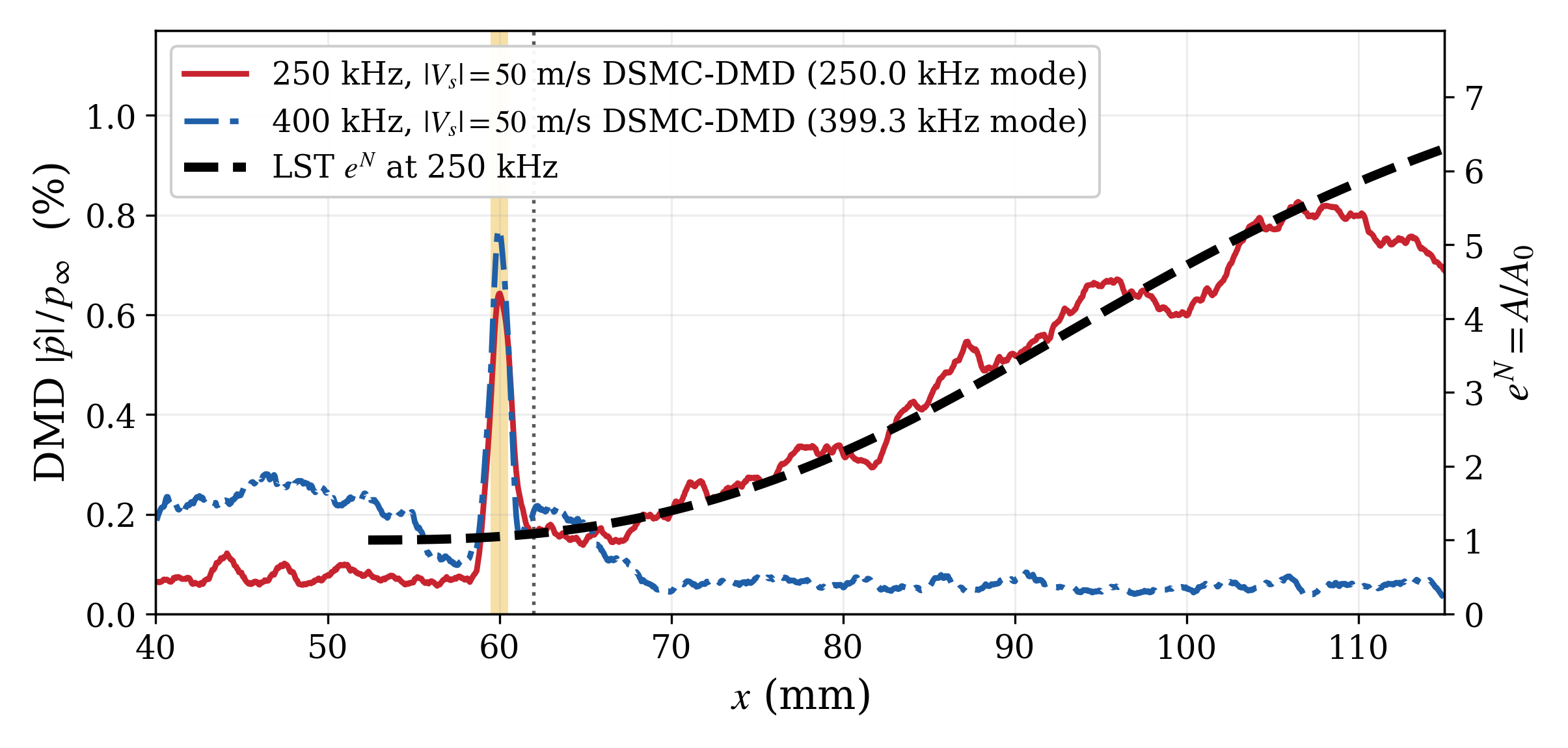}
\caption{Streamwise amplitude of the pressure DMD mode closest to the imposed AVS frequency, at $y=0.2$~mm, for the two $|V_s|=50$~m/s cases demonstrating that the $250$~kHz mode is amplified versus the decay of the  $400$~kHz mode in agreement with LST. DSMC-DMD amplitudes $|\hat{p}|/p_\infty$ are on the left ordinate and the LST amplification ratio $e^{N}$ at $250$~kHz on the right ordinate, tied to the $250$~kHz curve at $x=62$~mm (dotted line), downstream of the near-field peak over the AVS strip (yellow band).}
\label{fig:avs_amplification}
\end{figure}

A more quantitative view of the forced response is obtained from the streamwise amplitude of each pressure DMD mode, sampled at $y=0.2$~mm, near the wall, and expressed as a percentage of the freestream pressure.
The spatial amplification of the AVS-seeded 250~kHz wave can be quantified by the amplification ratio $e^N$,
\begin{equation}
    N(x, F) = \int_{x_0}^{x} -\alpha_i(x',y_M) \, dx'  = \ln\!\left(\frac{A(x)}{A_0}\right),
    \label{eq:N_factor}
\end{equation}
where \(y_M\) is the wall-normal location of the amplified disturbance wave and \(A_0 = A(x_0)\) is the amplitude at \(x_0\), the lower branch of the neutral curve beyond which the wave is amplified. Figure~\ref{fig:avs_amplification} shows this amplitude for the two $50$~m/s cases, with the forced yellow AVS strip location indicated,  together with the locally parallel LST amplification ratio $e^{N}$ at $250$~kHz. Although this is not the traditional {\it transition} $N$-factor, it can be seen that the $250$~kHz growth follows the LST $e^{N}$ trend, confirming that the AVS-seeded wave amplifies where linear theory predicts. The $400$~kHz mode carries content upstream of the strip, consistent with unforced instability there, peaks locally at the strip, and decays within about 10~mm downstream of it, consistent with its position outside the unstable band. These results show that the AVS interacts with the second-mode instabilities in a frequency-selective manner consistent with linear stability theory.

\begin{figure}[H]
\centering
\includegraphics[width=\columnwidth]{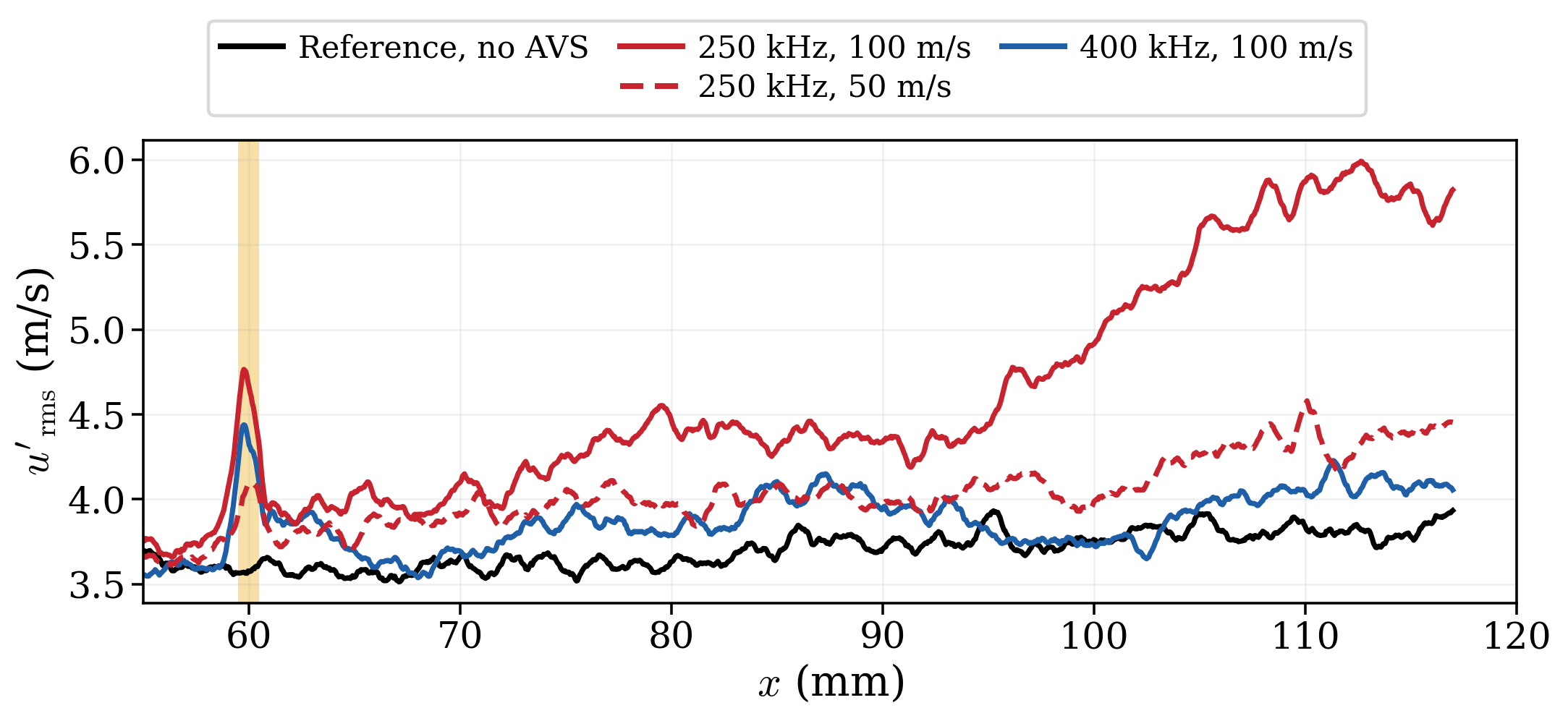}
\caption{Streamwise-velocity RMS along $x$ at $y\simeq0.3$~mm for the no-AVS reference, the $250$~kHz AVS cases at $|V_s|=50$ and 100~m/s, and the $400$~kHz case at 100~m/s. The yellow band marks the AVS strip location.
}
\label{fig:avs_rms_x}
\end{figure}

Figure~\ref{fig:avs_rms_x} shows the RMS perturbation amplitude, Eq.~\eqref{eq:rms}, of the streamwise velocity for the no-AVS reference, the $250$~kHz cases with a forcing velocity of $|V_s|=50$ and 100~m/s, and the $400$~kHz case at 100~m/s.  The AVS motion injects additional energy at its location, but downstream the two forcing frequencies behave differently. Downstream of the strip, the RMS of the  $400$~kHz case stays close to that of the no-AVS reference even at the larger forcing amplitude of $|V_s|=100$~m/s, whereas both $250$~kHz cases rise above it.   The flow is stable at $400$~kHz from the strip onward, so the forced  disturbance decays downstream. In contrast, the AVS seeds growing second-mode waves at
 $250$~kHz causing the disturbance RMS to grow downstream because the boundary layer is locally unstable at that frequency. The distinction between the two cases is more apparent for the larger AVS disturbance at 100~m/s. However, it can still be seen even at a value of $|V_s|=50$~m/s.

\begin{figure}[H]
\centering
\includegraphics[width=\columnwidth]{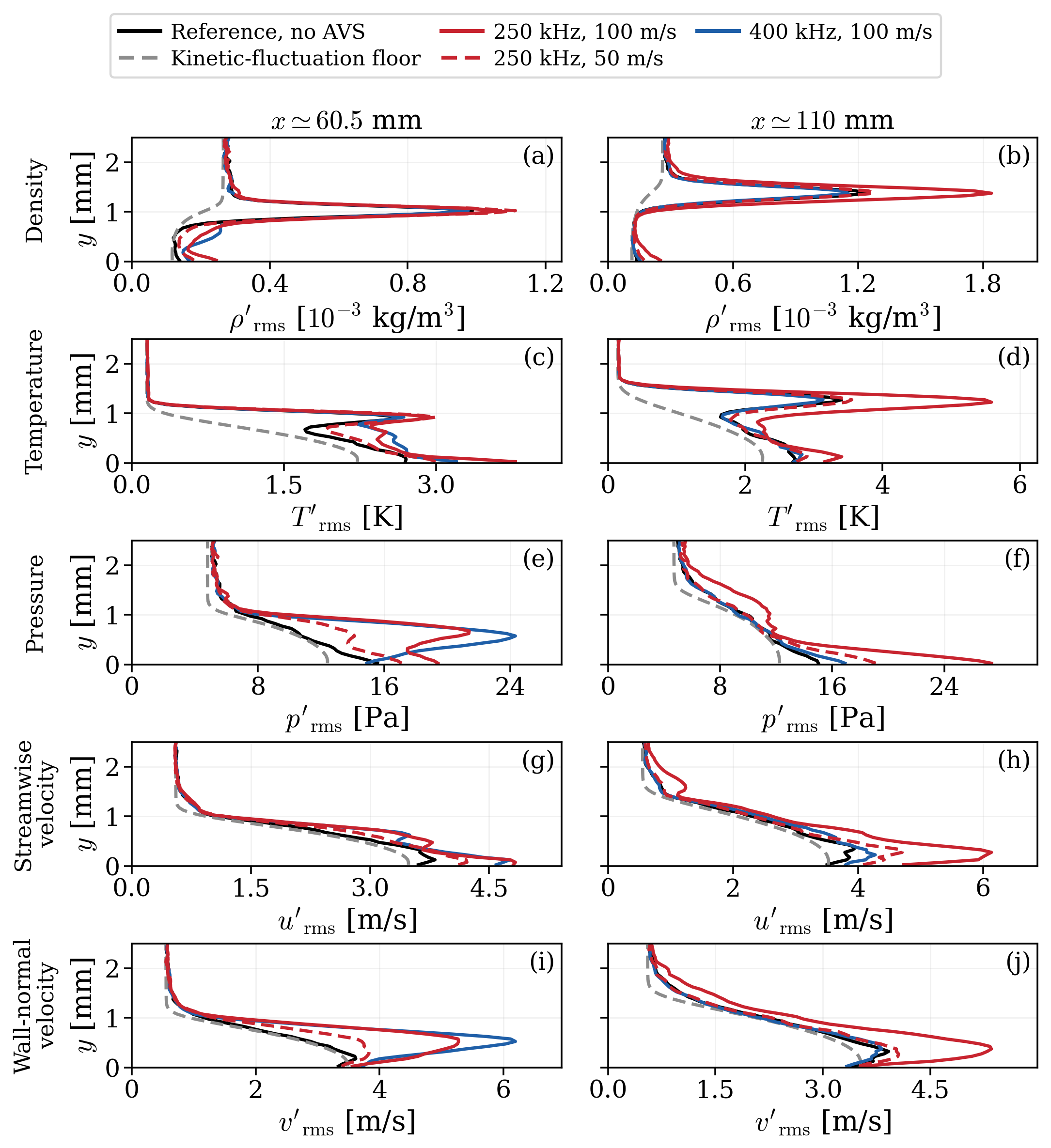}
\caption{Wall-normal RMS profiles of different flow variables at $x\simeq60.5$~mm (above the AVS) and $x\simeq110$~mm (downstream), for the no-AVS reference, the kinetic-fluctuation floor, the $250$~kHz AVS cases at $|V_s|=50$ and 100~m/s, and the $400$~kHz case at 100~m/s. Downstream, the $400$~kHz case stays close to the reference with residual differences, and the $250$~kHz cases show a larger excess.
}
\label{fig:avs_rms_y}
\end{figure}

Finally, Fig.~\ref{fig:avs_rms_y} provides a complementary cross-layer view of the wall-normal RMS profiles of selected flow variables just above the AVS ($x\simeq60.5$~mm) and downstream ($x\simeq110$~mm).  It can be seen in the figure that just above the AVS strip, both frequencies create a local disturbance that is greater than the no-AVS reference case.  However, further downstream, the $400$~kHz and reference RMS profiles stay close, with residual differences of a few percent over most of the layer and below 10 percent nearly everywhere, while the $250$~kHz profile remains elevated by its excited second-mode content even for the $|V_s|=50$~m/s case. At $|V_s|=100$~m/s the $250$~kHz profile peaks lie about 40 to 80 percent above the reference.
 The analytical kinetic-fluctuation floor discussed in Sec.~\ref{sec:dsmcnoise} is less than the reference and $400$~kHz profiles, since both still carry the second mode amplified from the kinetic fluctuations of the simulated gas.  Furthermore,  the $250$~kHz profile is even higher above that floor since at that frequency the AVS creates additional unsteady waves for both surface-velocity amplitudes.  Note that the  AVS thus adds growing fluctuations at the frequency that is amplified by the boundary layer.

The $50$~m/s cases demonstrate (Figs.~\ref{fig:avs_density_modes}--\ref{fig:avs_amplification}) that an AVS with a corresponding surface deflection of 31.8~$\mu$m is more than sufficient to excite and interact with the second mode at $250$~kHz. The wave seeded at $250$~kHz grows, whereas the disturbances seeded at the stable frequency of $400$~kHz decay downstream of the strip, in accordance with LST. The AVS therefore acts very similarly to the idealized periodic forcing of linear theory. It excites a second-mode wave only at its own frequency, and that wave is amplified where it is unstable, consistent with established continuum simulations and theory~\cite{egorov2006suctionblowing,liu2021wallperturbations}. The $100$~m/s cases confirm that the interaction remains frequency selective even at increased forcing strength (Figs.~\ref{fig:avs_rms_x} and~\ref{fig:avs_rms_y}) and that the time-averaged flow remains unaltered. The AVS therefore does not distort the base flow or modify its response. Just downstream of the strip ($x\simeq60.5$~mm, $y<1.5$~mm), the time-averaged pressure differs from the no-AVS reference by less than 0.25 percent and the density by less than 0.6 percent (Supplemental Material Sec.~\ref{sec:avsmeanflow}~\cite{SM}). A synthetic jet in a similar Mach~5.86 flat-plate boundary layer, in contrast, blows and sucks gas through a wall slot and generates additional modes through nonlinear interactions~\cite{zhuangActiveTransitionControl2023}. The AVS moves no mass through the wall and excites only the waves at its forcing frequency. By leaving the mean flow unchanged, it also leaves the growth of the waves at other frequencies undisturbed, which is itself a desirable quality for interactions with instabilities.

\section{Conclusions}
\label{sec:conclusions}

In this study, hypersonic boundary-layer instabilities over a flat plate were investigated using the particle-kinetic DSMC method coupled with data-driven DMD analyses.  Closed-form kinetic-theory expressions were derived to predict the amplitude of the DSMC fluctuations in the window-averaged snapshots.
It was observed that the DSMC-simulated flowfield naturally developed second Mack mode instabilities without any external seeding, {\em i.e.,} arising spontaneously from the kinetic fluctuations of the simulated gas and rising above the predicted kinetic-fluctuation floor.  The
frequencies of these perturbations were captured via local PSD analysis and found to fall within the range predicted by linear stability theory. Notably, near the surface at $x=80$~mm the perturbations lie within the locally unstable frequency band, whereas along the boundary-layer edge the most energetic frequency at each location lies close to the upper neutral curve, indicating that these instabilities are amplified and accumulate as they convect downstream. For a more thorough examination, the time-accurate unsteady DSMC flowfield snapshots were analyzed using DMD to extract global spatial structures of the instability waves at the frequencies identified in the PSD spectra.  DMD extracts the individual waves that make up the second-mode wave packets.  The DMD amplitude envelopes show that for each of the captured modes, growth occurs within their respective unstable regions predicted by linear stability theory.  This is consistent with the DMD modes being the decomposed structures of second-mode wave packets.

With the demonstration of the ability of  combined DSMC and DMD approaches to model second Mack mode instabilities, we investigated the possible interactions of an AVS to selectively excite hypersonic boundary layer disturbances in the 200 to 500~kHz range.
An efficient approach to model the interaction of the AVS with the boundary layer flow was developed by modifying the reflected particle velocity distribution function, instead of actually moving the surface.
An AVS oscillating at a frequency within the locally unstable region was shown to  interact with boundary-layer instabilities, introducing perturbations that persist and amplify in the predicted growth zone, following the LST $e^N$ trend. In contrast, an AVS at a stable-region frequency generated perturbations that decayed within about 10~mm of the strip without sustained propagation.

Overall, the present study shows that the DSMC method, integrated with data-driven techniques, offers a robust approach for investigating unsteady hypersonic boundary layer flows in the context of second Mack mode instabilities. The continued improvement in DSMC algorithms enables us to model a number density of \(1.2\!\times\!10^{24}\,\text{m}^{-3}\) and a Reynolds number of $Re_L=1.4\times10^{6}$, very high for a particle kinetic method.
While such conditions correspond to a low Reynolds number from a continuum perspective, to the best of our knowledge, they are the highest number densities a particle method has reached for an atmospheric boundary-layer flow. Furthermore, the implementation of a simple AVS provides a kinetic-level model of a vibrating surface,  along the lines of the moving-surface DSMC implementations of Rader, Gallis and Torczynski~\cite{raderDSMCMovingBoundary2011}.

Questions regarding the role of receptivity and free stream fluctuations still remain important to the hypersonics community and will be considered in our future work.
Addressing this problem by DNS is rather challenging due to the presence of strong shocks and shock--boundary-layer and entropy-layer interactions,  near-tip rarefaction and nonequilibrium effects. The leading edge region is also where  rarefaction effects such as slip velocity and temperature jump are strong. Velocity slip has been shown to stabilize second-mode waves and to destabilize first-mode waves, while temperature jump destabilizes the second mode~\cite{Ou2021SlightlyRarefied}. While slip velocity and temperature jump boundary conditions
may be incorporated in DNS, a DSMC approach to the present problem could also be employed, whereby rarefied gas phenomena are captured naturally, the flow evolves from collisional length scales, and slip velocity and temperature jump emerge from gas molecule--surface interactions rather than being imposed as flow boundary conditions.

\begin{acknowledgments}
This work was supported by the Office of Naval Research under Grant No. N00014-23-1-2839, "Kinetic Treatment of Sources and Mechanisms that Drive Unsteady, Shock-dominated Flow Instability" (Program Officer: Dr. Eric Marineau). The authors gratefully acknowledge Professor Vassilis Theofilis for his useful discussions and suggestions regarding this work, in particular on the linear stability analysis. Computational resources were provided by the Texas Advanced Computing Center (TACC) at the University of Texas through the Frontera supercomputer (Project No. CTS23002), funded by the National Science Foundation (NSF), and by Purdue University's Anvil supercomputer through ACCESS allocation PHY250034.
\end{acknowledgments}

\newpage

\bibliography{bibliography}

% ---------------- Supplemental Material ----------------
\clearpage
\setcounter{section}{0}
\setcounter{figure}{0}
\setcounter{table}{0}
\setcounter{equation}{0}
\renewcommand{\thesection}{S\arabic{section}}
\renewcommand{\thefigure}{S\arabic{figure}}
\renewcommand{\thetable}{S\arabic{table}}
\renewcommand{\theequation}{S\arabic{equation}}
\renewcommand{\theHsection}{S\arabic{section}}
\renewcommand{\theHfigure}{S\arabic{figure}}
\renewcommand{\theHtable}{S\arabic{table}}
\renewcommand{\theHequation}{S\arabic{equation}}

\begin{center}
{\large\bfseries Supplemental Material for: A Kinetic Study of the Interactions of an Acoustic Vibrating Surface with Hypersonic Boundary-Layer Instabilities}\\[1ex]
Mert Senkardesler, Irmak T. Karpuzcu, and Deborah A. Levin\\
\textit{University of Illinois, Urbana-Champaign, IL 61801, USA}
\end{center}

This Supplemental Material gives the particle-count convergence tests of the direct simulation Monte Carlo (DSMC) simulations in Sec.~\ref{sec:numconv} and the compressible similarity boundary layer, with its comparison to DSMC, in Sec.~\ref{sec:blas}. Section~\ref{sec:2ndmoderes} gives the linear stability theory (LST), the point power spectral densities (PSDs) and the dynamic mode decomposition (DMD) method, Sec.~\ref{sec:avsmeanflow} compares the time-averaged flow with and without the acoustic vibrating surface (AVS), and Sec.~\ref{sec:derivation} derives the speed distribution of molecules reflected from the AVS. Its references are included in the reference list above.

\section{DSMC Numerical Convergence Tests \label{sec:numconv}}

The three DSMC numerical parameter test cases that have been considered are outlined in Table~\ref{tab:cases},
all using the same physical models and DSMC numerical parameters given in Table~\ref{tab:params} of the paper, except for \(\FNUM\). The Baseline case
is considered the primary case and will be compared with simulations performed with fewer and higher number of
particles, ``Coarse 1.2B case'' and ``Fine 20B case'', respectively. Here \(\FNUM\) is the DSMC particle-weight ratio or
the number of real gas particles represented by each DSMC simulation or computational particle.  The three cases differ only in \(\FNUM\), and hence in the particle population (\(N_p\), the particles per \(\lambda_\infty^2\) and the mean number per sampling cell, \(\bar N_{p,\mathrm{cell}}\)) and in the number of adaptive collision cells, \(N_\mathrm{coll}\) (Table~\ref{tab:cases}). These cases were chosen because the number of simulation
particles sets the variance of the macroparameters predicted by a DSMC simulation, such as number density, temperature, velocity,
and pressure, while their mean should not depend on it. It is well known that DSMC follows Poisson statistics as one samples or
averages over particles during some chosen timespan to obtain an average stochastic flow quantity with a variance or
RMS.  The RMS fluctuation level therefore scales with the inverse square root of the effective sample count.
It can be reduced either by increasing the number of DSMC particles in the sampled volume or by sampling that
volume over a longer steady-state time interval.

\begin{table}[H]
\centering
\caption{Three test cases with varying computational particle counts.}
\label{tab:cases}
\begin{tabular}{lccccc}
\toprule
Case & \(\FNUM\) & \(N_p\) & Particles per \(\lambda_\infty^2\) & \(\bar N_{p,\mathrm{cell}}\) & $N_\mathrm{coll}$ \\
\midrule
Coarse 1.2B & \(2.88\times 10^{11}\) & $1.2\times 10^9$ & 1.575 & $8.38\times 10^3$ & $\approx 1.4\times 10^{8}$ \\
Baseline & \(8.525\times 10^{10}\) & $4.1\times 10^9$ & 5.321 & $2.84\times 10^4$ & $\approx 4.7\times 10^{8}$ \\
Fine 20B & \(1.728\times 10^{10}\) & $2.0\times 10^{10}$ & 26.25 & $1.40\times 10^5$ & $\approx 1.1\times 10^{9}$ \\
\bottomrule
\end{tabular}
\end{table}

\begin{figure}[H]
\centering
\includegraphics[width=0.95\linewidth]{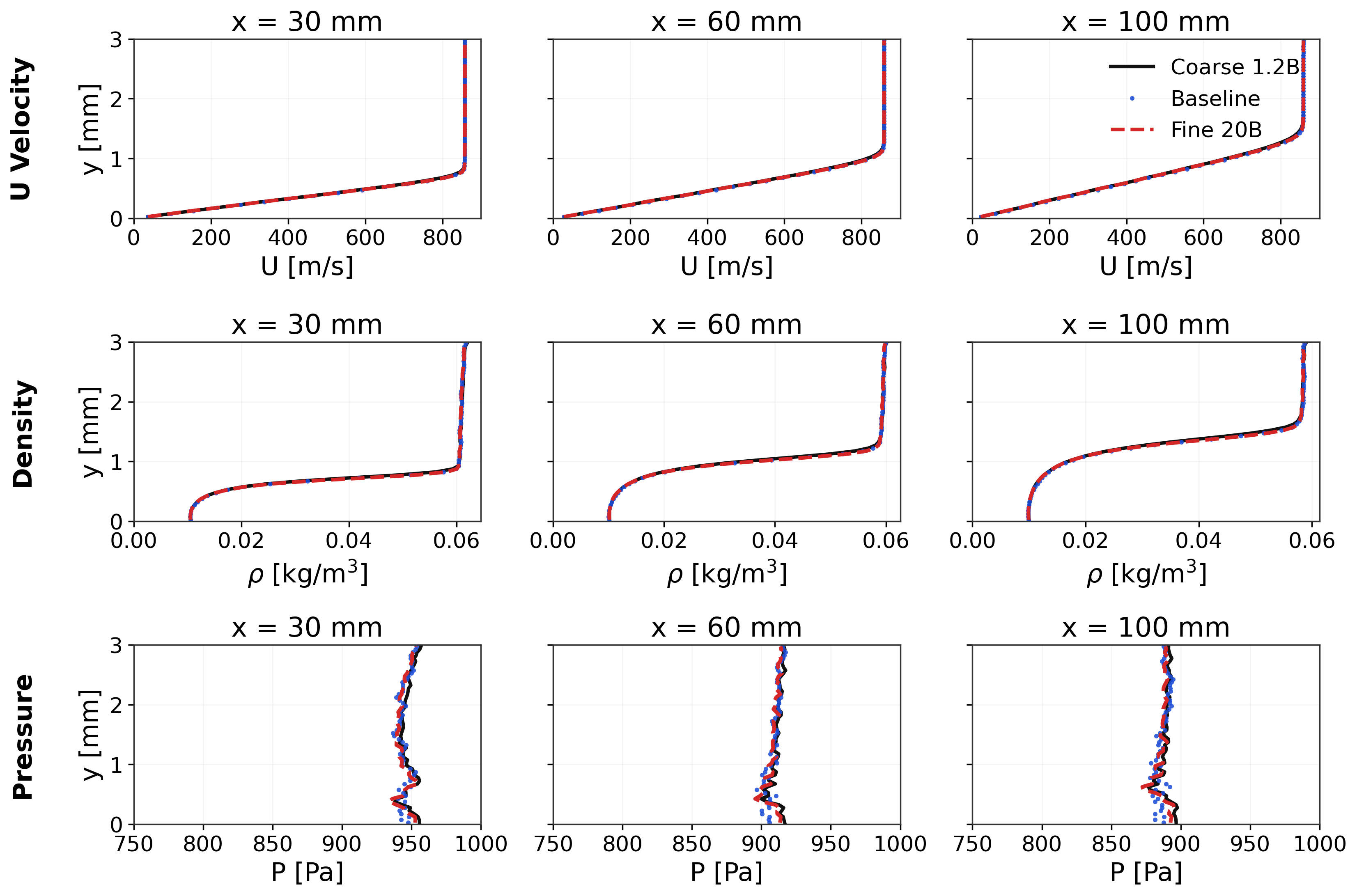}
\caption{Mean boundary-layer profiles for the Coarse 1.2B, Baseline (blue dots), and Fine 20B cases at \(x=30\), 60, and 100 mm.}
\label{fig:meanprofiles}
\end{figure}

Starting with the effect of the number of computational particles on the mean boundary layer, Figure~\ref{fig:meanprofiles}
compares the Coarse 1.2B, Baseline, and Fine 20B profiles at three representative streamwise locations. It can be
seen that the typical deviation is less than 1\% indicating that the mean structure remains
sufficiently close to use the Baseline particle count ($4.1\times10^{9}$) as the baseline case for all simulations presented in this work.

\begin{figure}[H]
\centering
\includegraphics[width=0.95\linewidth]{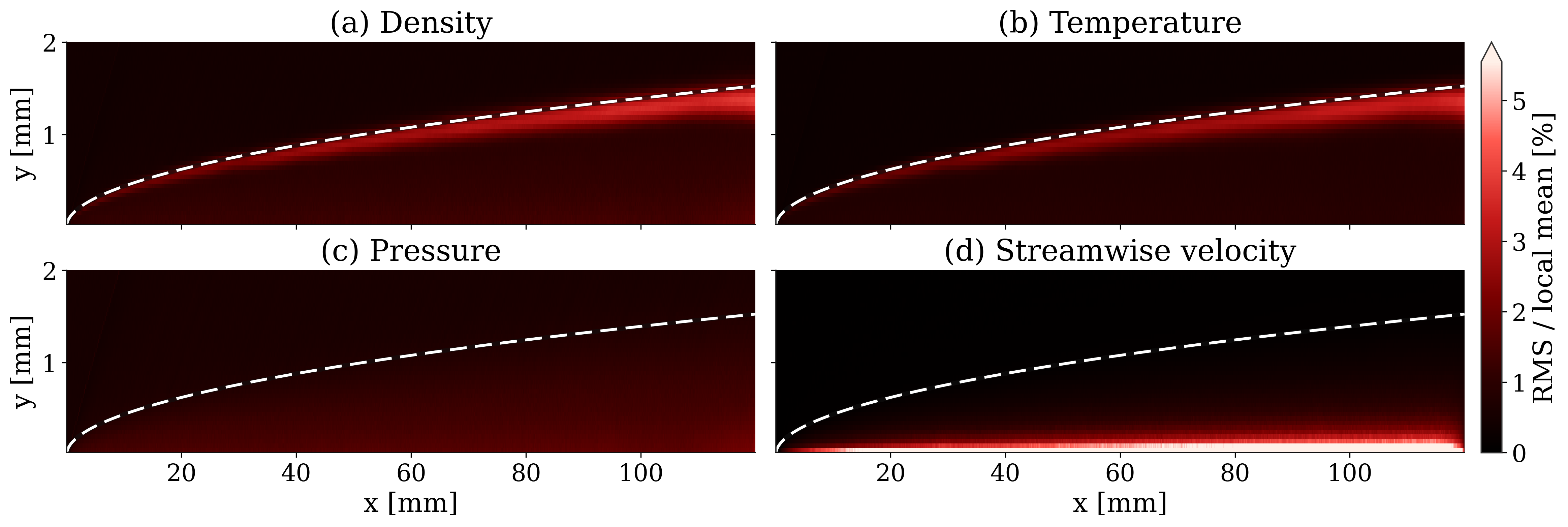}
\caption{Baseline RMS-percentage fields over \(0\le y\le2\) mm. Each subfigure shows \(100q'_{\mathrm{rms,lin}}/\bar q\) for the indicated variable, with density plotted as physical mass density \(\rho\); for streamwise velocity the denominator is \(|\bar u|\). The white dashed curve marks the analytical compressible-Blasius boundary-layer height \(\delta^{\mathrm{sim}}_{99}(x)\) computed from the same Baseline edge-state reference used in the mean-profile verification. All four subfigures use the same percentage color scale so the relative fluctuation levels can be compared directly across \(\rho\), \(T\), \(p\), and \(u\).}
\label{fig:rmsfields}
\end{figure}

Table~\ref{tab:broadband} gives the freestream probe values and Figure~\ref{fig:rmsfields} shows the spatial distribution of the DSMC fluctuations, {\em i.e.,} the mean-subtracted fluctuation \(q'_t=q_t-\bar q\) given by Eqs.~\eqref{eq:fluctuation} and \eqref{eq:rms}.
These RMS values come directly from the DSMC data using Eq.~\eqref{eq:rms} of the paper over a bandwidth of about
 10 kHz to 10 MHz. Because the stored
DMD snapshot spacing is \(\Delta t_s=50\) ns, the fully resolved fluctuation band in these records runs
from the record-scale resolution limit of about 10 kHz up to the Nyquist limit of 10 MHz.
To obtain RMS values of density, the original DSMC post-processed signal is converted to
mass density from the sampled particle count per cell, the particle weight \(\FNUM\) and the molecular mass. The freestream value at \((x,y)=(1,2)\) mm is \(\rho_\infty=5.58\times10^{-2}\) kg/m\(^3\).
 The DSMC RMS values show the expected particle-count
convergence.  That is,  increasing the number of particles in the sampled volume reduces the fluctuation amplitude, and, the
decrease follows the anticipated inverse-square-root scaling with effective particle count. For context, freestream pitot-pressure fluctuations for a hypersonic quiet tunnel are of order 0.01\% of the mean in quiet operation and about 3\% in noisy operation~\cite{nicotra2026freestream}. In the same tunnel, boundary-layer RMS values were calculated from pressure spectra integrated over 50 to 400~kHz~\cite{nicotra2026freestream}. The DSMC RMS values shown in Table~\ref{tab:broadband} are obtained from sampled data of a much wider dynamic range, 10~kHz to 10~MHz. Restricted to the same 50 to 400~kHz band, the freestream \(p'_{\mathrm{rms}}/\bar p\) of the Fine 20B case is 0.058\%.

\begin{table}[H]
\centering
\caption{DSMC broadband freestream RMS values in one sampling cell at \((x,y)=(1,2)\) mm.\textsuperscript{a,b}}
\label{tab:broadband}
\begin{tabular}{lcccc}
\toprule
Quantity & \(\bar q\) & Coarse 1.2B RMS & Baseline RMS & Fine 20B RMS\\
\midrule
\(p'_{\mathrm{rms}}\) [Pa] & 828.3 & $7.898\;(0.95\%)$ & $4.486\;(0.54\%)$ & $1.969\;(0.24\%)$\\
\(\rho'_{\mathrm{rms}}\) [kg/m\(^3\)] & $5.58\times10^{-2}$ & $4.40\times10^{-4}\;(0.79\%)$ & $2.49\times10^{-4}\;(0.45\%)$ & $1.12\times10^{-4}\;(0.20\%)$\\
\(u'_{\mathrm{rms}}\) [m/s] & 860.62 & $1.015\;(0.12\%)$ & $0.532\;(0.06\%)$ & $0.246\;(0.03\%)$\\
\(T'_{\mathrm{rms}}\) [K] & 50.02 & $0.277\;(0.55\%)$ & $0.150\;(0.30\%)$ & $0.067\;(0.13\%)$\\
\bottomrule
\end{tabular}
\vspace{0.5em}

\footnotesize
\textsuperscript{a} Analysis performed over 700,000--800,000 DSMC time steps.\\
\textsuperscript{b} Numbers in parentheses represent RMS percentage fluctuation or \(\%(q_{\mathrm{rms}})=100 q_{\mathrm{rms}}/\bar q\).
\end{table}

Figure~\ref{fig:rmsfields} computed using Eq.~\eqref{eq:rms} of the paper gives the flowfield RMS distribution for density, temperature,
pressure, and streamwise velocity. Comparison of the four subfigures shows that the same unsteady DSMC field does
not project with the same spatial structure for every macroscopic variable. For example, Figures~\ref{fig:rmsfields}(a)
and~\ref{fig:rmsfields}(b) show that the largest relative density and temperature fluctuations form in a band near the
outer part of the boundary layer, close to the boundary-layer edge. However, the pressure field
in Figure~\ref{fig:rmsfields}(c) is organized differently, with its strongest amplitudes concentrated inside the boundary
layer and shifted closer to the surface and Figure~\ref{fig:rmsfields}(d) shows that the streamwise-velocity RMS is even
more wall-focused, with the strongest relative amplitudes in the near-wall portion of the layer. The table and the figure
alone, however, do not allow us to separate the DSMC RMS fluctuations into physical versus numerical.
For that we need formulations of the theoretical DSMC error as a function of both the number of computational particles as well as the degree of sampling independence.  This is discussed in the paper (see Eqs.~\eqref{eq:rho_theory}--\eqref{eq:epsilon}).

\section{Hypersonic Flat-Plate Boundary Layer \label{sec:blas}}

To study the behavior of boundary layer instabilities and to determine specific locations for DSMC probes, the boundary layer general structure must be known. In this section we start with the well known compressible Blasius solution. Using the boundary layer thickness as the characteristic length, the Knudsen number $Kn$ is given by \cite{chambre2017flow}:
\begin{equation}
    Kn = \frac{\lambda_w}{\delta_{99}} = \sqrt{\frac{\gamma \pi}{2}}\, \frac{M_w}{\text{Re}_{\delta,w}}
\end{equation}
where \(M_w = U_e/a_w\) and \(\text{Re}_{\delta,w} = \rho_w U_e \delta_{99}/\mu_w\) are based on the wall density, viscosity and speed of sound. At \(x = 10\) mm from the leading edge the DSMC simulation gives \(\lambda_w = 6.07\)~\(\mu\)m, \(\delta_{99} = 0.460\) mm, and \(Kn = 0.013\). This value indicates that the flow near the leading edge falls within the slip flow regime ($0.01 \leq Kn \leq 0.1$) meaning that
 gas molecules exhibit a finite tangential velocity relative to the surface. The relaxation of the no-slip condition alters the boundary layer structure by reducing the shear stress exerted by the body on the flow, {\em i.e.,} velocity slip leads to weaker near-wall velocity and temperature gradients. This results in lower second-mode growth rates and a larger first-mode unstable region compared to predictions from no-slip, continuum approaches~\cite{Wang2024SlipReceptivity,Ou2021SlightlyRarefied}.

 The magnitude of the slip velocity is given by:
\begin{equation}
    u_{\text{slip}} = \frac{2 - \sigma_v}{\sigma_v} \lambda \frac{\partial u}{\partial y} + \frac{3}{4} \frac{\mu}{\rho T} \frac{\partial T}{\partial x} \label{eq:slip}
\end{equation}
where \( \sigma_v \) is the tangential momentum accommodation coefficient (taken as 1.0 in the present DSMC simulations), \( \lambda \) is the mean free path, \( u \) is the velocity in the streamwise direction, and \( y \) is the wall-normal direction.
Equation~\eqref{eq:slip} can be used to modify the no-slip boundary condition in continuum simulations such as the compressible flat plate flow solutions. It is not imposed in DSMC, where slip follows from the gas-surface interaction model, here diffuse reflection with full accommodation.

\begin{figure}[H]
    \centering
    \includegraphics[width=0.5\linewidth]{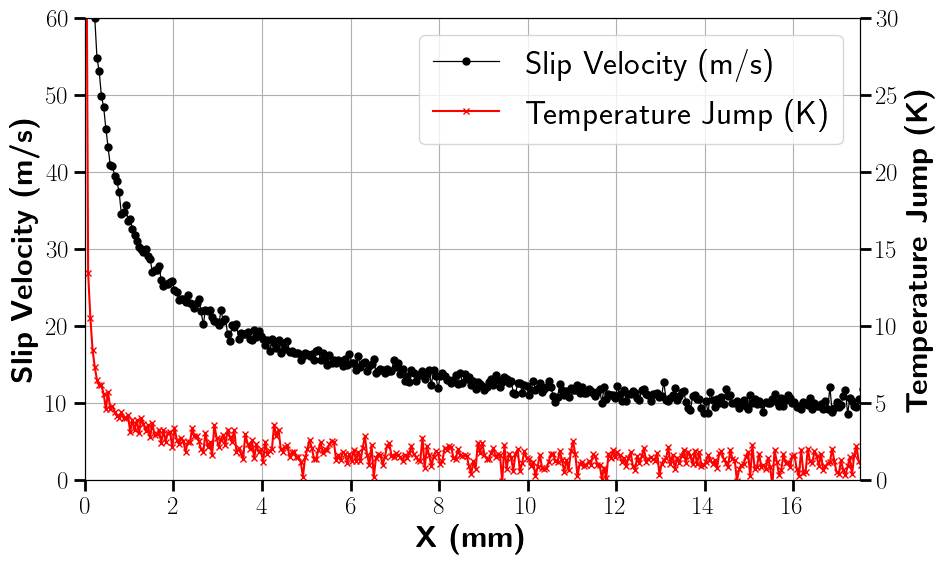}
    \caption{Slip velocity and temperature jump profiles along the surface from the DSMC solution.}
    \label{fig:BLdictate}
\end{figure}

It was found that velocity slip is relevant in the near vicinity of the leading edge, within the first 15~mm. Figure~\ref{fig:BLdictate} shows profiles of slip velocity and temperature jump obtained from the DSMC simulation. It can be seen in the figure that near the leading edge, slip velocities reach approximately 60~m/s, accompanied by steep pressure gradients, but, these effects diminish downstream, with slip velocity falling to around 10~m/s by 15~mm. It can also be seen that the temperature jump in the developed region of the simulation is only on the order of 1~K and therefore will be neglected. The compressible similarity solution that follows is therefore obtained with a no-slip wall.

The theoretical solution for compressible boundary-layer flow over a flat plate can now be obtained using the Crocco method \cite{CroccoBL}. This theory assumes the gas is calorically perfect with a constant specific heat capacity while accounting for temperature-dependent viscosity and thermal conductivity. For compressible flow over a flat plate, a self-similar solution within the boundary layer can be obtained through an appropriate transformation of the independent variables. In the transformed plane \((\xi, \eta)\), the velocity profile becomes independent of \(\xi\) and can be expressed as \(u = u(\eta)\).
The independent variables in the transformed space \((\xi, \eta)\) relate to those in the physical space \((x, y)\) by:
\begin{equation}
    \xi = \int_{0}^{x} \rho_e u_e \mu_e \, dx\quad \eta = \frac{u_e}{\sqrt{2\xi}} \int_{0}^{y} \rho \, dy
    \label{eq:xi}
\end{equation}
The governing equations in the transformed plane \((\xi, \eta)\) are given by:
\begin{equation}
    (C f'')' + f f'' = 0
    \label{eq:momentum}
\end{equation}
\begin{equation}
    \left( \frac{C}{Pr} g' \right)' + f g' + C \frac{u_e^2}{h_e} \left( f'' \right)^2 = 0
    \label{eq:energy}
\end{equation}
where \(C = \frac{\rho\mu}{\rho_e \mu_e}\), \(\rho\) and \(\mu\) are the density and dynamic viscosity, respectively, and the subscript \(e\) denotes values at the edge of the boundary layer. The Prandtl number \(Pr = \frac{\mu c_p}{k}\), where \(c_p\) is the specific heat capacity at constant pressure, and \(k\) is the thermal conductivity. The power law viscosity model is used in this study, given by $\mu = \mu_{\text{ref}} \left( \frac{T}{T_0} \right)^{n}$
where \(\mu_{\text{ref}}\) is the reference viscosity, \(T\) is the temperature, \(T_0\) is the reference temperature, and \(n\) is the exponent. In this study, \(n = 0.75\) for molecular nitrogen, matching the variable hard sphere gas model of the DSMC simulations.
The solution to the boundary-layer equations requires appropriate boundary conditions at the wall (\(\eta = 0\)). For flow with velocity slip at the wall, these boundary conditions are:
\begin{equation}
    f(0) = 0, \quad f'(0) = u_{\text{slip}}/U_e^*, \quad g(0) = g_w \quad \text{for an isothermal wall}
\end{equation}
Because \(u_{\text{slip}}\) was negligible, it was set to zero.
By solving these transformed equations with the specified boundary conditions, we obtain the variations of velocity and enthalpy across the boundary layer, leading to the self-similar solution for the compressible boundary layer over a flat plate. The dependent variables of Eqs.~\eqref{eq:momentum} and \eqref{eq:energy} can be obtained from the transformed variables as:
\begin{equation}
    u = f' {U_e^*} \quad \text{and} \quad h = {h_e} g
    \label{eq:velocity}
\end{equation}
where \( U_e^* \) denotes the dimensional velocity at the boundary layer edge.

\begin{figure}[H]
\begin{subfigure}[b]{0.49\linewidth}
\includegraphics[width=\linewidth]{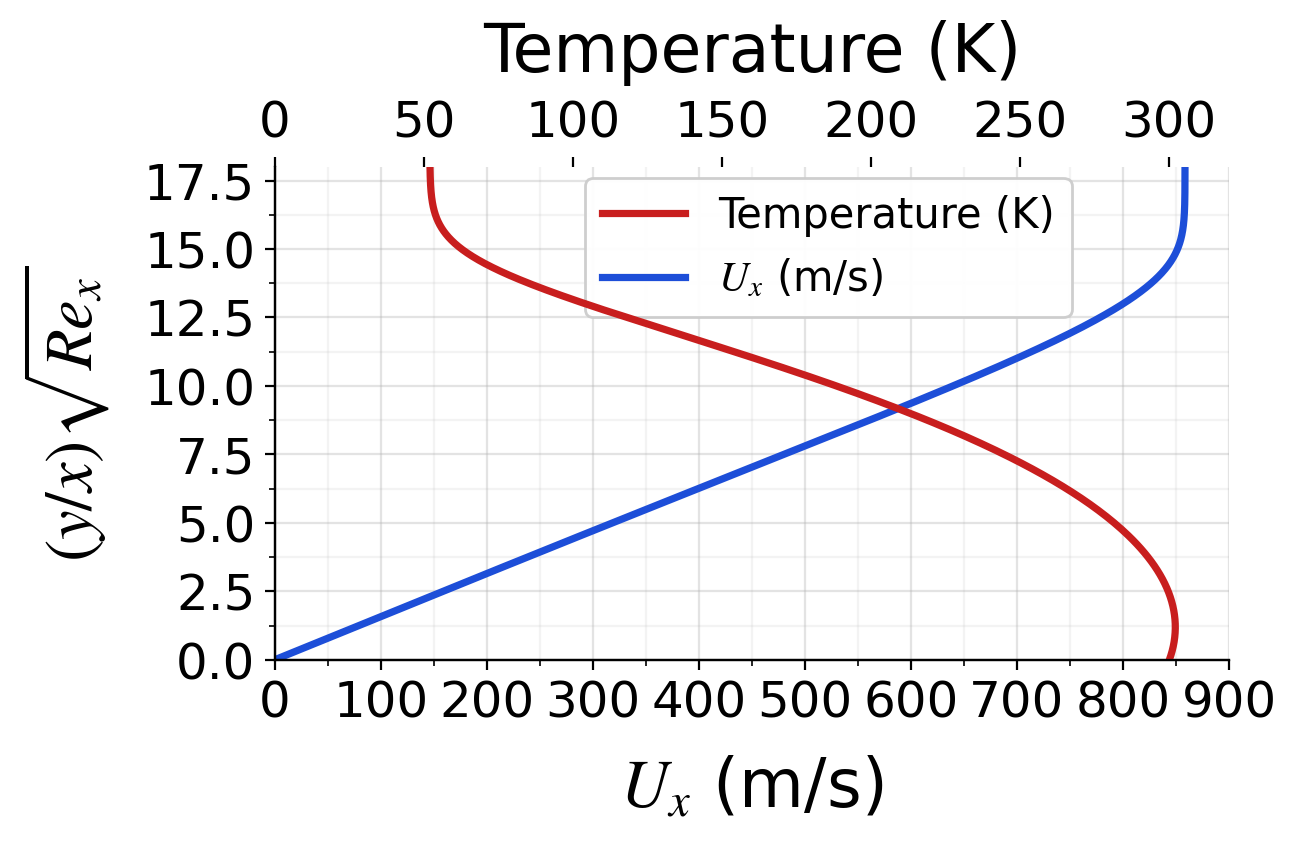}
\caption{Temperature and streamwise velocity profiles versus the self-similar height.}
\label{subfig:blasius_profiles}
\end{subfigure}\hfill
\begin{subfigure}[b]{0.47\linewidth}
\includegraphics[width=\linewidth]{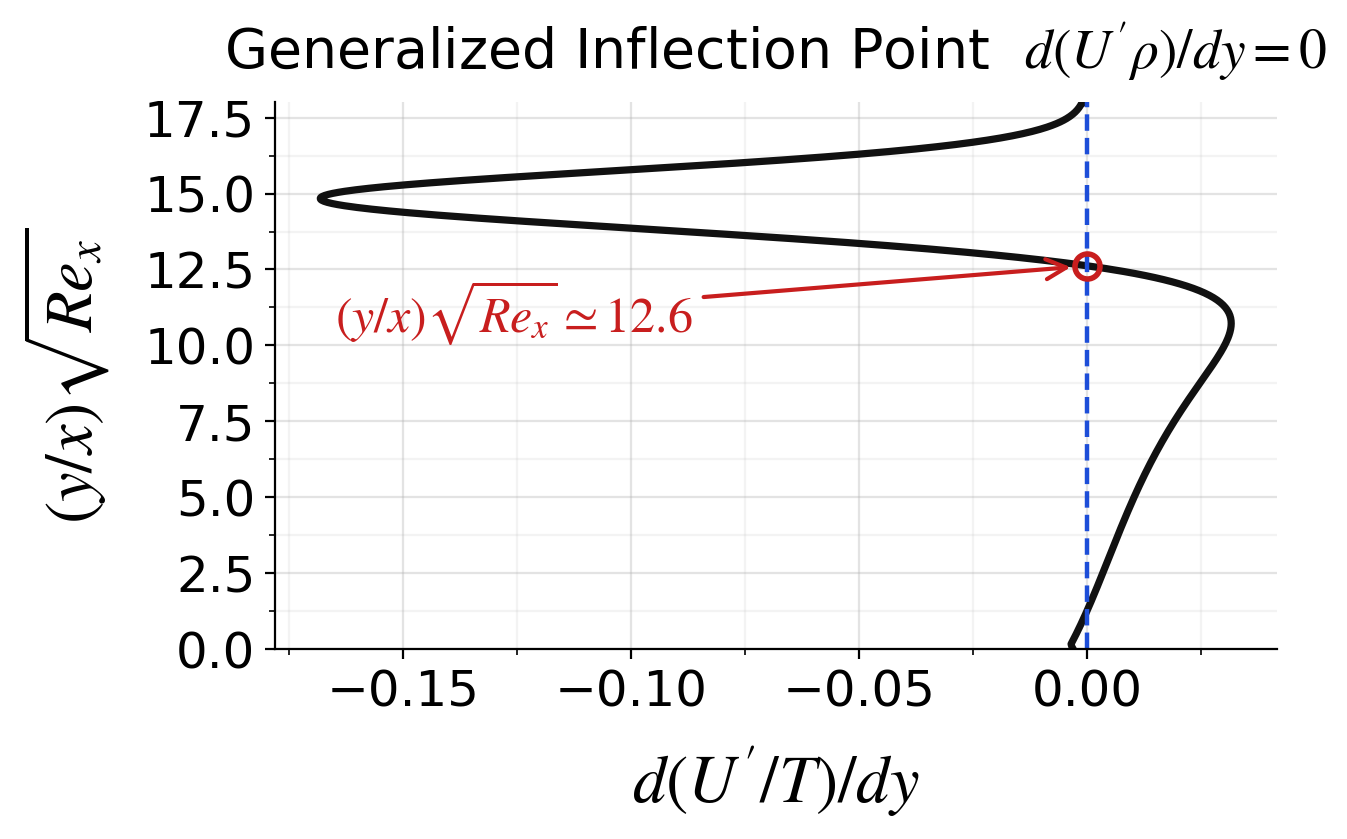}
\caption{Generalized inflection point for Mach 5.85 at $y\sqrt{Re_x}/x \approx 12.6$.}
\label{subfig:inflection_point}
\end{subfigure}
\caption{Compressible Blasius boundary-layer profiles and generalized inflection point. (a) Temperature and streamwise velocity profiles versus the self-similar height. (b) Generalized inflection point from the compressible Blasius equations for Mach 5.85 at $y\sqrt{Re_x}/x \approx 12.6$.}
\label{fig:combined_subfigures}
\end{figure}

The compressible Blasius boundary layer solution provides the baseline boundary-layer profiles, expressed in terms of the similarity coordinate $\eta$ and the functions $f(\eta)$ and $g(\eta)$. The obtained solution is shown in Fig.~\ref{fig:combined_subfigures}(a), which illustrates the non-monotonic temperature profile characteristic of hypersonic boundary layers. The solution also exhibits a generalized inflection point, defined as the location where
\begin{equation}
\left. \frac{d}{dy} \left( \rho \frac{dU}{dy} \right) \right|_{y=y_g} = \quad \frac{\rho_e^2 u_e^2}{2\mu_e x}\,\frac{f''' g - 2f'' g'}{g^{4}} = 0
\label{eq13}
\end{equation}
Here \(\rho=\rho_e/g\) across the layer, so the generalized inflection point is where \(f'''g-2f''g'=0\).

This point indicates the necessary condition for the existence of inviscid instabilities, as it satisfies the Lees--Lin generalized inflection-point condition, the compressible form of Rayleigh's theorem. As shown in Fig.~\ref{fig:combined_subfigures}(b), the generalized inflection point occurs at a height of $y\sqrt{Re_x}/x \simeq 12.6$, while the boundary-layer edge ($u/U_e = 0.99$) lies at $y\sqrt{Re_x}/x \simeq 14.9$, placing the inflection point at $0.85\,\delta_{99}$, well inside the layer. Here $Re_x$ is formed on edge properties, as required by the similarity identity $(y/x)\sqrt{Re_x} = \sqrt{2}\int_0^{\eta}(T/T_e)\,\mathrm{d}\eta'$.

\begin{figure}[H]
\begin{subfigure}[b]{0.49\linewidth}
\includegraphics[width=\linewidth]{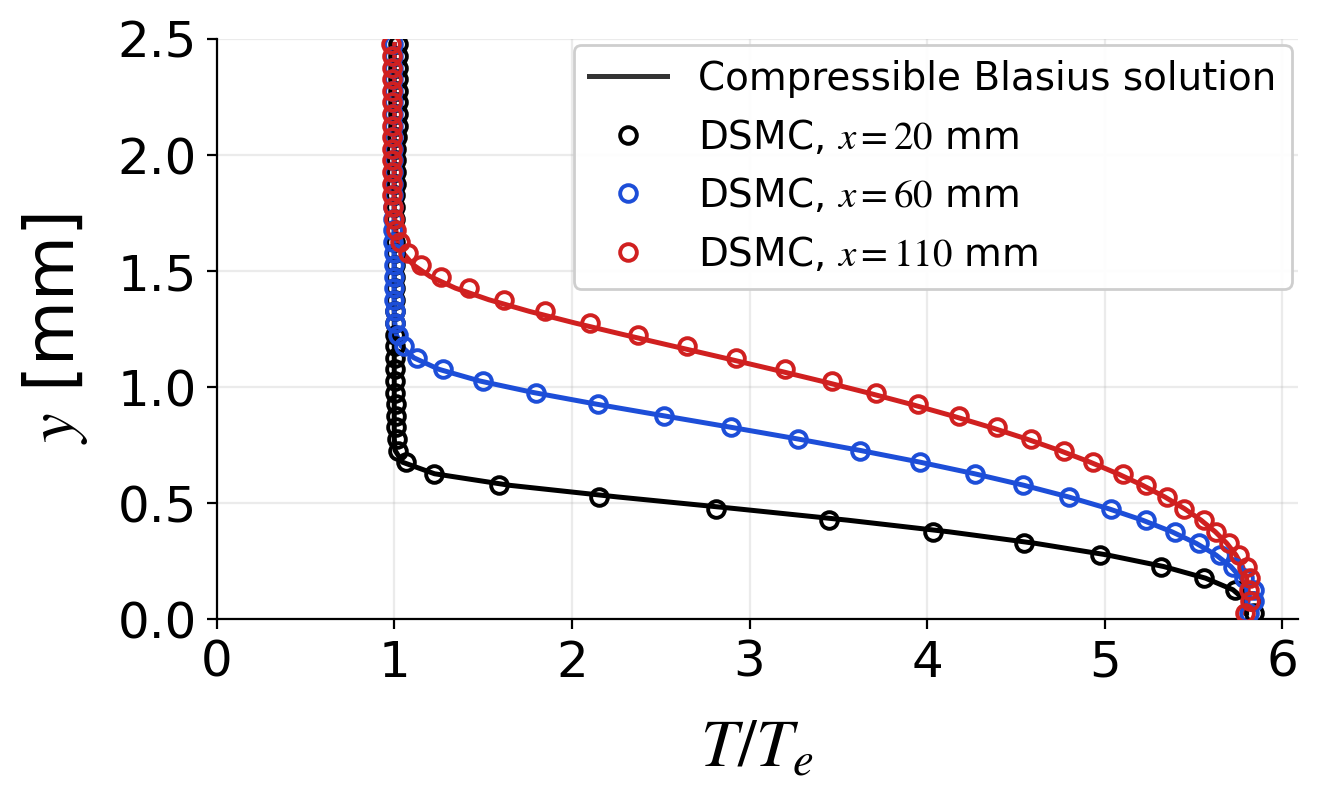}
\caption{Temperature profiles.}
\label{subfig:blasius_temp}
\end{subfigure}
\begin{subfigure}[b]{0.49\linewidth}
\includegraphics[width=\linewidth]{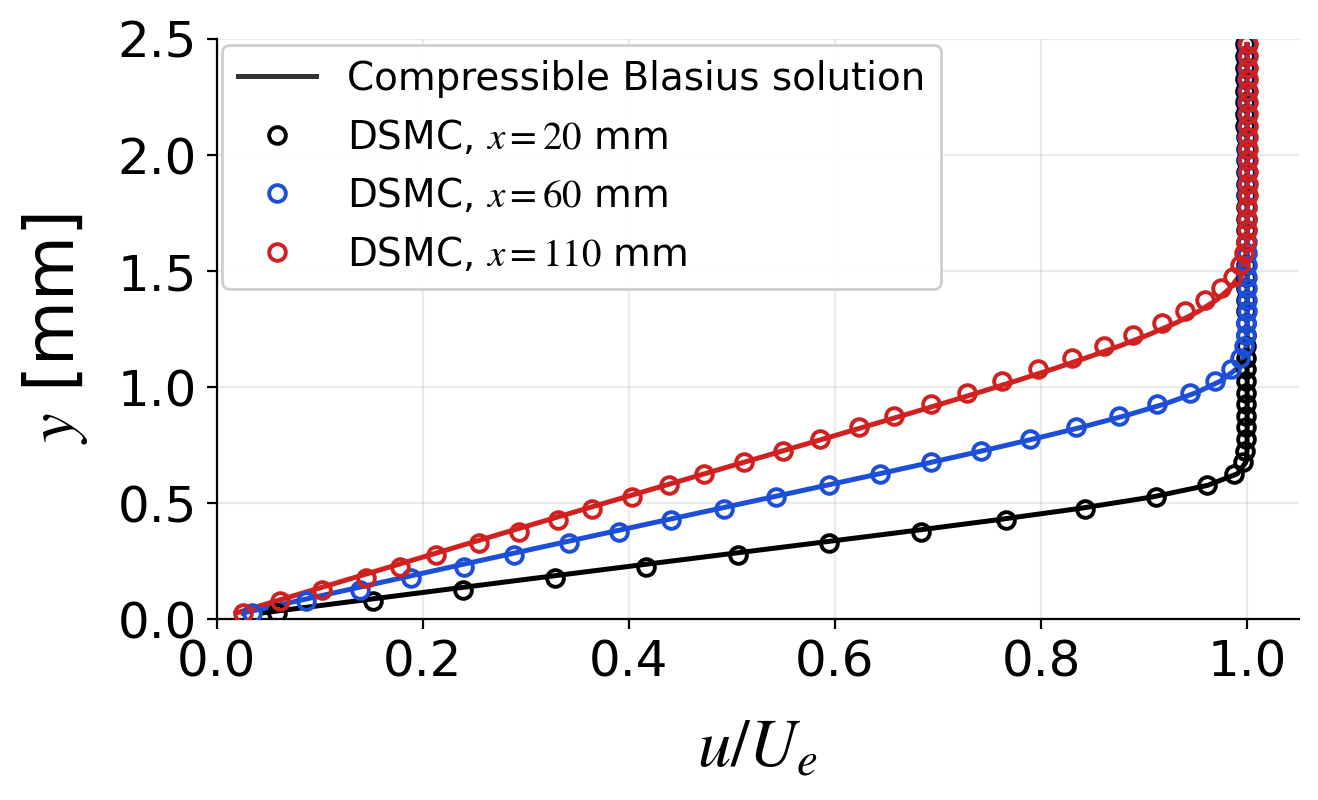}
\caption{Velocity profiles.}
\label{subfig:blasius_vel}
\end{subfigure}
\caption{Comparison of DSMC results with the compressible Blasius solution at Mach 5.85, edge temperature 52\,K, edge velocity 858\,m/s, Prandtl number 0.72, and wall temperature 300\,K, at $x=20$, $60$ and $110$ mm: (a) temperature profiles and (b) velocity profiles.}
\label{fig:boundaryTOTAL}
\end{figure}

\begin{figure}[H]
    \centering
    \includegraphics[width=0.6\linewidth]{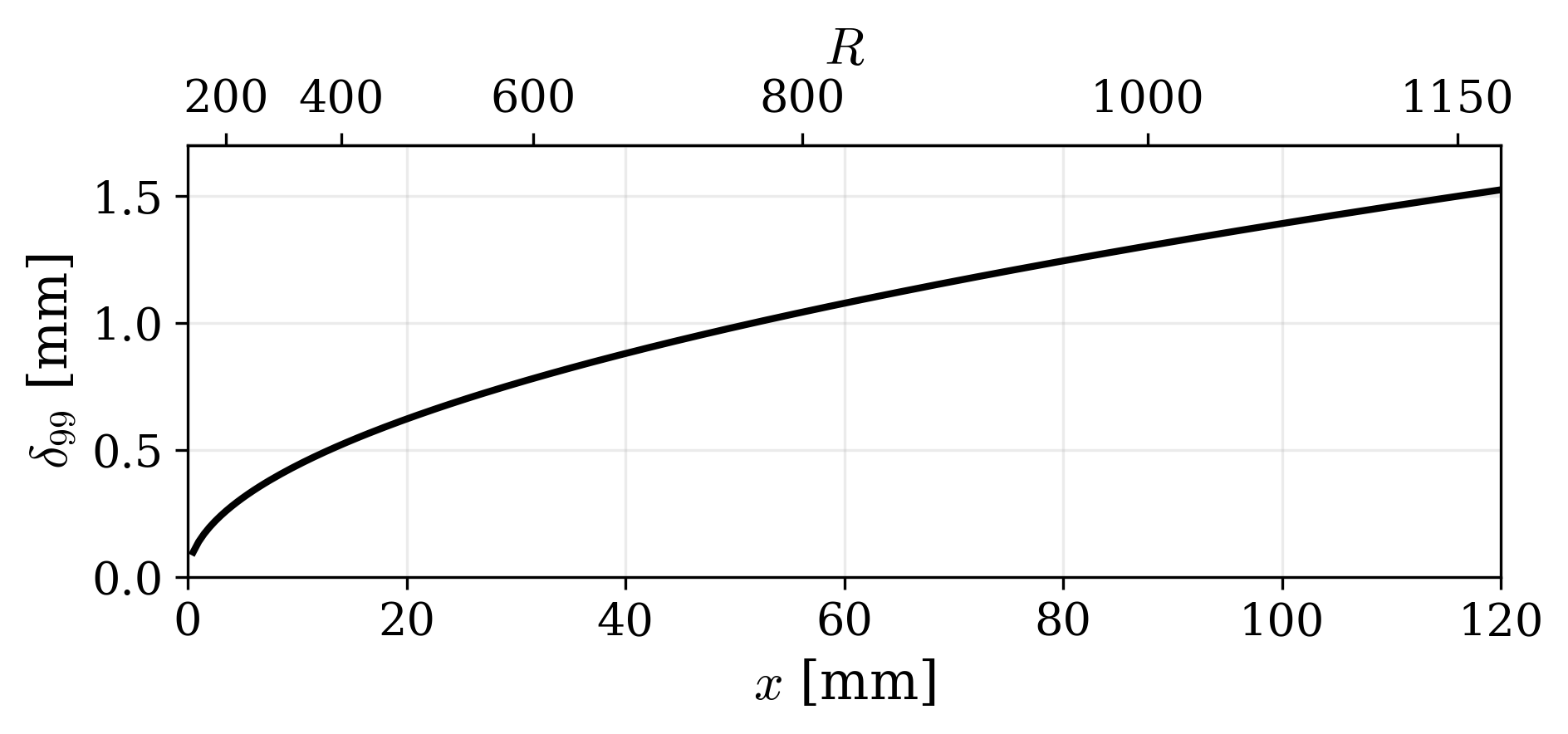}
    \caption{Boundary-layer thickness profile obtained from the compressible Blasius equations as a function of distance and dimensionless streamwise coordinate for the conditions specified in Figure~\ref{fig:boundaryTOTAL}.}
    \label{fig:flat plate_profile}
\end{figure}

Comparison of velocity and temperature profiles predicted by the compressible Blasius solution with DSMC results presented in Fig.~\ref{fig:boundaryTOTAL} shows agreement, confirming the use of the compressible Blasius equations to help select meaningful probe locations for DSMC simulations at this Reynolds number.
More specifically, the compressible Blasius equations are used to determine the boundary-layer thickness $\delta_{99}$ along the streamwise direction $x$,
as shown in Fig.~\ref{fig:flat plate_profile}. The $\delta_{99}$ values will be used to efficiently place virtual probes in the DSMC simulations near the edge of the boundary layer where second-mode instabilities exhibit peak amplitudes.

\section{DSMC Resolution of  Second-mode Instabilities\label{sec:2ndmoderes}}

\subsection{Linear Stability Characterization of the Second Mack Mode \label{sec:secondmacmode}}

Linear stability theory for  2D instabilities postulates that the evolution of unsteady macroparameters can be written as
\begin{equation}
    \label{eq:normalModeX}
 {u_t}(x, y) \;=\; \hat{u}(y) \,
    \exp \!\Big[ \int \big(- \alpha_i + i \alpha_r\ \big)dx -
    \,i\,\;\omega\,t \Big],
\end{equation}
where the complex streamwise wavenumber \(\alpha = \alpha_r + i\alpha_i\) governs the spatial characteristics of disturbances in hypersonic boundary layers. The real component, \(\alpha_r\), determines the spatial wavelength of the disturbance through \(\lambda = 2\pi/\alpha_r\), defining the streamwise periodicity of the perturbation waves. The imaginary component, \(\alpha_i\), controls the spatial growth or decay rate, measured by \(-\alpha_i\), the logarithmic streamwise derivative of the disturbance amplitude.
A positive \(-\alpha_i > 0\) indicates exponential growth of the perturbation amplitude as it propagates downstream, characteristic of convectively unstable modes, while \(-\alpha_i < 0\) signifies damping, and \(-\alpha_i = 0\) marks neutral stability.

The neutral curves and growth rates follow from local linear stability theory~\cite{mackBoundaryLayerLinearStability}. At each streamwise location the base flow is the compressible similarity profile of Sec.~\ref{sec:blas}, taken as locally parallel. The linearized compressible disturbance equations for two-dimensional waves are then solved as a spatial eigenvalue problem, with a real frequency and the complex wavenumber \(\alpha\) as the eigenvalue. For a given frequency, the second-mode identity location \(R_{\min}\) is the most upstream location at which the pressure eigenfunction shows the single phase reversal of the second mode.

The stability calculation uses the same edge and wall state as the DSMC simulation: a self-similar compressible boundary layer of molecular nitrogen (\(\gamma = 1.4\)) at edge Mach number \(M_e = 5.85\), edge static temperature \(T_e = \SI{52}{\kelvin}\), and an isothermal wall at \(T_w = \SI{300}{\kelvin}\), giving a wall-to-edge temperature ratio \(T_w/T_e = 5.77\). The transport closure is a constant Prandtl number \(Pr = 0.72\) together with a power-law viscosity \(\mu = \mu_{\mathrm{ref}}(T/T_0)^{n}\) with \(n = 0.75\).
 The neutral stability curves used in this work are verified by zero neutral decay rates.
To enable comparison between DSMC results and the neutral stability curves, dimensional variables are converted into dimensionless forms consistent with boundary-layer stability theory. The dimensionless streamwise coordinate is defined as:
\begin{equation}
R = \sqrt{\frac{U_e^* x^*}{\nu^*}} = \sqrt{Re}, \quad L^* = \sqrt{\frac{\nu^* x^*}{U_e^*}},
\label{eq:ReynoldsAndLength}
\end{equation}
where \(U_e^*\) is the local velocity at the boundary-layer edge, \(x^*\) is the dimensional distance from the leading edge, \(\nu^*\) is the kinematic viscosity, and \(Re\) is the Reynolds number based on \(x^*\). The dimensionless angular frequency \(\omega\) is related to the dimensional cyclic frequency \(\omega^*\) (in kHz) by:
\begin{equation}
\omega = \frac{2 \pi \omega^* L^*}{ U_e^*}, \quad  F=\frac{\omega}{R}
\label{eq:omega}
\end{equation}
incorporating the local length scale \(L^*\) and edge velocity \(U_e^*\).
These transformations enable direct comparisons between the theoretical stability curves (expressed in
$\omega $ and $R$) and DSMC-derived results (in kHz and mm), providing a consistent framework to assess the onset and growth of second-mode instabilities.

Figure~\ref{fig:edge_psd} of the paper shows the resulting second-mode neutral curve for the present conditions, overlaid on the DSMC spectra.
Disturbances with frequencies and streamwise positions falling within the enclosed unstable regions experience spatial amplification.  The outer boundary of each enclosed region corresponds to \(-\alpha_i = 0\), marking neutral waves.  Inside these regions, where \(-\alpha_i > 0\), perturbations exhibit exponential spatial growth as they propagate downstream, defining a convectively unstable zone until they exit, where \(-\alpha_i \leq 0\) indicates damping or neutral stability.

\subsection{Spectral Characterization of Boundary Layer Instabilities} \label{sec:baselineLocalPSD}
With the boundary layer spatial features defined in Sec.~\ref{sec:blas}
and the  dimensional representation of neutral stability curves we analyze the spectral content of instabilities predicted by the DSMC kinetic approach.
Each signal is linearly detrended and analyzed over the statistically steady window from 700,000 to 800,000 timesteps, using the full-field DSMC snapshots stored every 50 timesteps (a sampling frequency of 20~MHz, 2001 snapshots). Each PSD is a single Hann-windowed periodogram of the full 100~$\mu$s record, as in Sec.~\ref{sec:flowInstabilities} of the paper, which gives a frequency bin spacing of 10~kHz.

\begin{figure}[H]
    \centering
    \begin{subfigure}[b]{0.48\textwidth}
        \includegraphics[width=\linewidth]{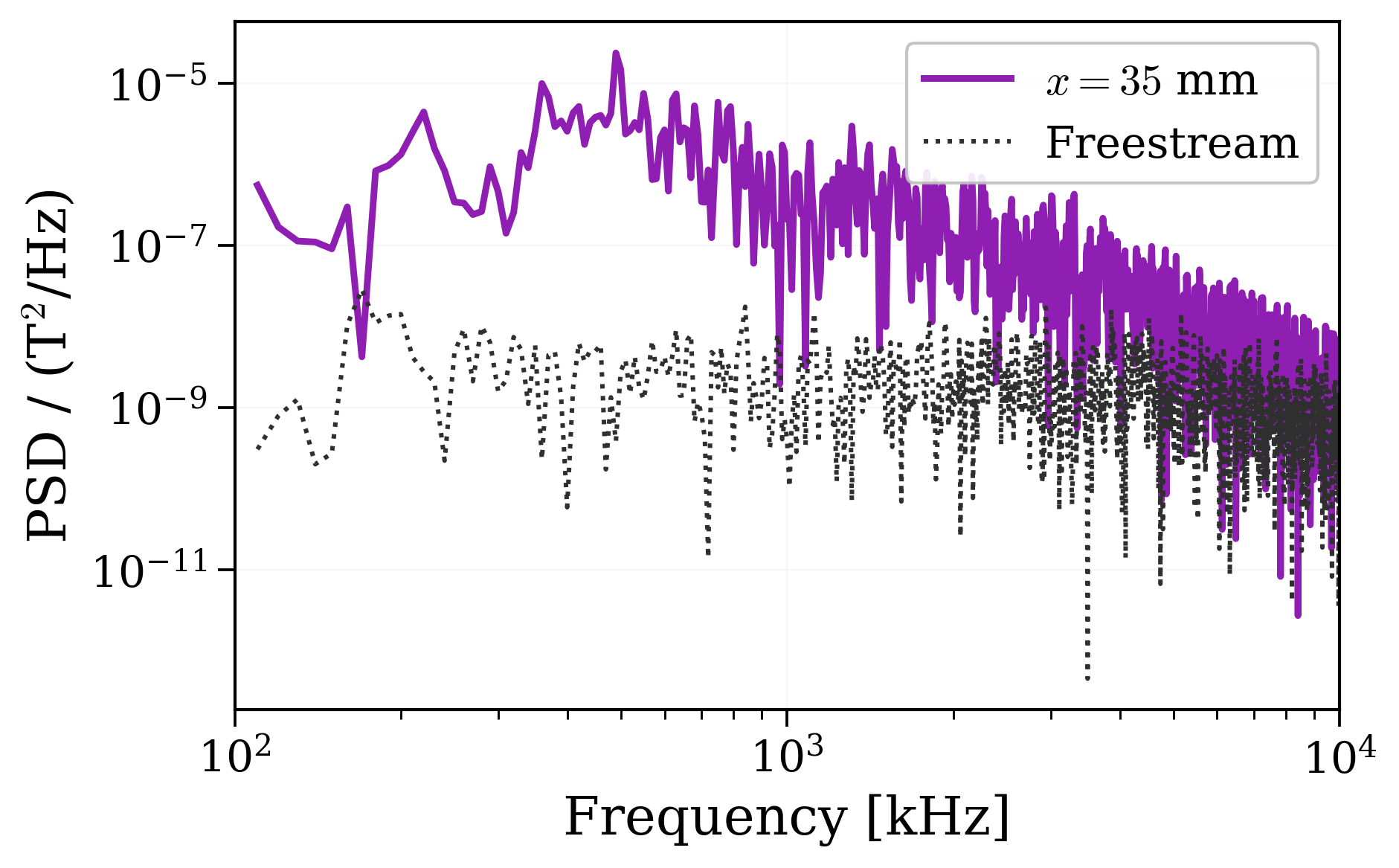}
        \caption{PSD at $x=35$\,mm.}
        \label{subfig:psd_spectrum_35}
    \end{subfigure}
    \begin{subfigure}[b]{0.48\textwidth}
        \includegraphics[width=\linewidth]{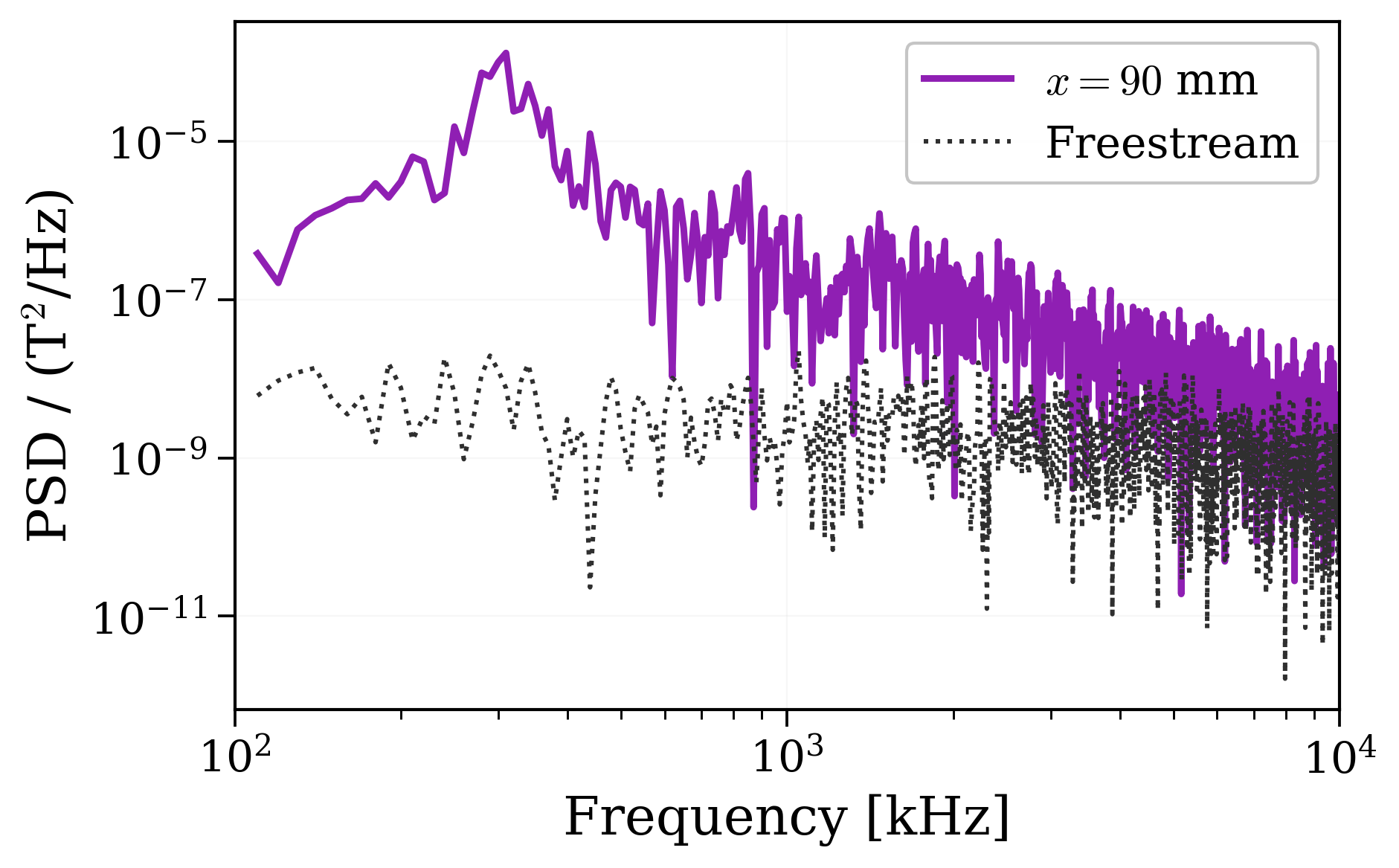}
        \caption{PSD at $x=90$\,mm.}
        \label{subfig:psd_spectrum_90}
    \end{subfigure}
    \caption{Temperature PSD. (a,b) Point PSD at $x=35$ and $90$~mm compared with the freestream, and well-defined second-mode peaks rise far above the flat freestream level.}
    \label{fig:PROBEDPSD}
\end{figure}

The statistically steady DSMC field is analyzed via PSD, as shown in Fig.~\ref{fig:PROBEDPSD}(a,b).  The first observation is that well-defined spectral peaks are seen at both axial locations, which is even more clear when compared to the
 flat spectral response of the freestream
that  is relatively free of disturbances. The point spectra shown, from the boundary layer, are taken at $(x, y) = (35, 0.75)$ and $(90, 1.23)$~mm, at $0.9\,\delta_{99}$.  Secondly, the spectral peaks from these two different locations are at different frequencies.  PSD analysis reveals distinct frequencies that evolve with streamwise coordinate: at \(x=\SI{35}{\milli\meter}\), a pronounced peak emerges at approximately \SI{490}{\kilo\hertz} [Fig.~\ref{fig:PROBEDPSD}(a)], while by \(x=\SI{90}{\milli\meter}\) the dominant peak shifts to \SI{310}{\kilo\hertz} [Fig.~\ref{fig:PROBEDPSD}(b)].

\subsection{Dynamic Mode Decomposition} \label{sec:dmdmethod}
\begin{figure}[H]
    \centering
    \includegraphics[width=0.9\linewidth]{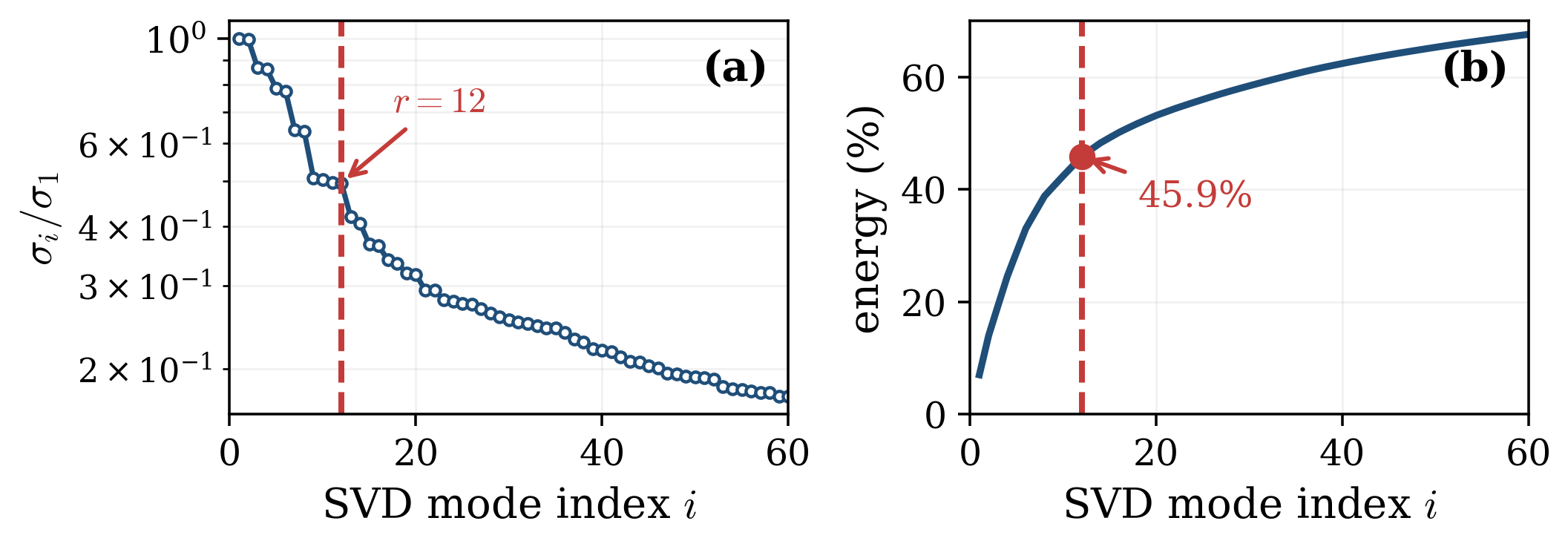}
    \caption{Singular-value spectrum of a density snapshot ensemble. (a) Normalized singular values $\sigma_i/\sigma_1$ versus SVD mode index, showing a knee near the retained rank $r=12$ (dashed line). (b) Cumulative captured snapshot energy, with the first 12 modes accounting for a cumulative energy of about 46\%.
    }
    \label{fig:svd}
\end{figure}

The DMD of the Baseline case uses the window snapshots from 750,000 to 800,000 timesteps, which gives 1001 snapshots at the 50-timestep spacing. The DMD field spans \(x=47\)--\(117\)~mm and \(y=0\)--\(2.1\)~mm, and each variable is decomposed separately with rank 12. For density, the 12 retained modes hold about 46\% of the snapshot energy (Fig.~\ref{fig:svd}). The AVS cases of Sec.~\ref{sec:PM_model} of the paper use the same window, with rank 14 on a DMD field spanning \(x=40\)--\(115\)~mm and \(y=0\)--\(2.5\)~mm.

When it comes to the DMD analyses, to quantitatively determine the wavelength of captured waves in the DMD modes, the amplitude is extracted along the boundary-layer edge in the streamwise direction to provide a one-dimensional signal, $u(x)$, representing the spatial oscillation of that mode \cite{senkardeslerDSMCStudyHypersonic2026}. To find the dominant spatial frequency, a Fast Fourier Transform (FFT) is applied to this signal \(u(x)\) thereby decomposing the signal from the spatial domain into the wavenumber domain, {\em i.e.,}
\begin{equation}
    \hat{u}(k_x) = \mathcal{F}\{u(x)\} = \int u(x) e^{-i k_x x} dx
    \label{eq:FFT_wavelength}
\end{equation}
where \(k_x\) is the streamwise wavenumber.   The resulting power spectrum, \(|\hat{u}(k_x)|^2\), exhibits a distinct peak at the dominant wavenumber, \(k_{x, \text{dom}}\). The wavelength \(\lambda\) of the disturbance and the dimensional phase speed \(c^*\) of the traveling wave then follow from this dominant wavenumber and the DMD-identified temporal frequency \(\omega^*\) as
\begin{equation}
    c^* = \lambda\,\omega^*,\qquad \text{where}\quad
    \lambda = \frac{2\pi}{k_{x, \text{dom}}}.
    \label{eq:wavelength_from_k}
\end{equation}

\begin{figure}[H]
    \centering
    \includegraphics[width=0.85\linewidth]{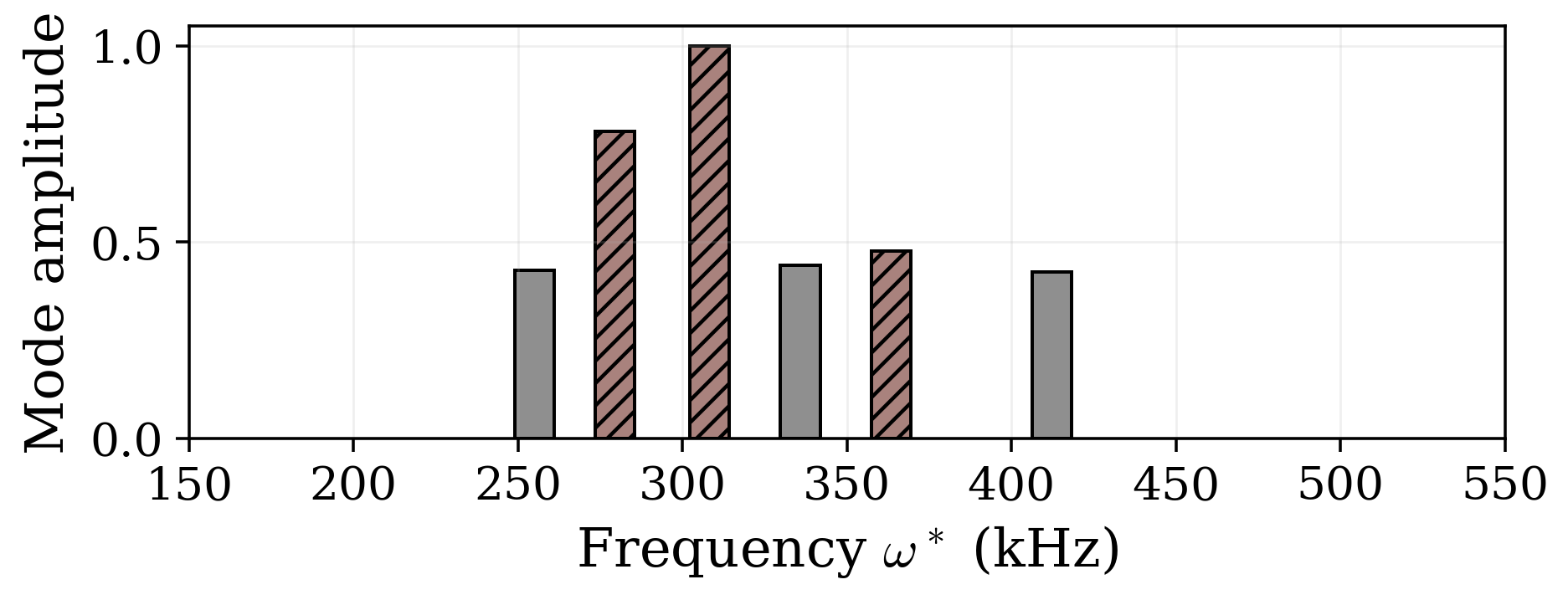}
    \caption{Frequency spectrum of the Baseline density DMD modes, with amplitudes normalized to the most dominant mode and the three most energetic modes indicated by crosshatching.}
    \label{fig:DMD_modes_freq}
\end{figure}

Figure~\ref{fig:DMD_modes_freq} shows the spectrum of the most energetic modes normalized by the most dominant mode of the DSMC predicted flow over a flat plate.  It can be seen that the dominant modes cluster between 250 and 420~kHz, consistent with the frequency range of second-mode instabilities identified in the PSD analysis [Fig.~\ref{fig:edge_psd} of the paper].  We retained  12  modes in the DMD analysis and due to complex-conjugate pairing, this yields 6 distinct frequencies, a number sufficient to resolve the dominant instability mechanisms. This rank is guided by the singular-value spectrum of the snapshot ensemble (Fig.~\ref{fig:svd}), which falls steeply over the first dozen modes and then flattens into a slowly-decaying tail.   Truncating at this knee retains the coherent instability structures while discarding the tail.

\raggedbottom
\section{Effect of the AVS on the Mean Flowfield}\label{sec:avsmeanflow}

\begin{figure}[H]
\centering
\includegraphics[width=\linewidth]{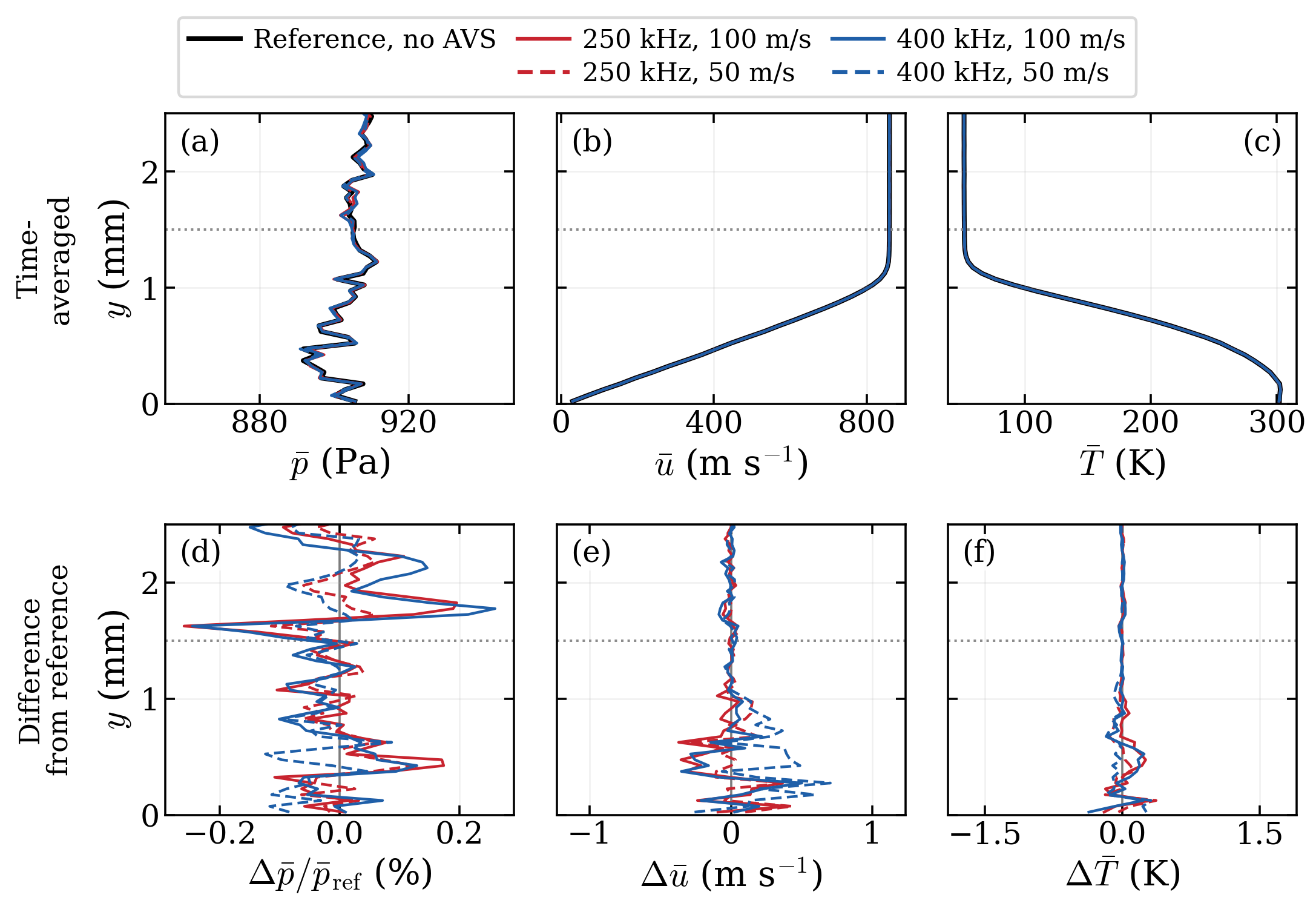}
\caption{Time-averaged flow at \(x=65\)~mm, steps 700\,000 to 800\,000, for the no-AVS reference and the
four AVS cases. (a)--(c) Wall-normal profiles of pressure, streamwise velocity and temperature. (d)--(f) Their
differences from the reference, relative for pressure and absolute for velocity and temperature.
The horizontal dotted line marks \(y=1.5\)~mm.}
\label{fig:avsmeanflow}
\end{figure}

Section~\ref{sec:PM_model} of the main text states that the AVS leaves the time-averaged flow
unaltered. Figure~\ref{fig:avsmeanflow} shows the evidence at \(x=65\)~mm, 4.5~mm downstream of the strip. The means
are taken over steps 700\,000 to 800\,000 in the cell column nearest that location.

The profiles of the four AVS cases and the reference coincide [Fig.~\ref{fig:avsmeanflow}(a)--(c)].
Just downstream of the strip, at \(x\simeq60.5\)~mm and \(y<1.5\)~mm, the mean pressure differs from the reference by
less than 0.25 percent, the density by less than 0.6 percent, the streamwise velocity by less than 1.1~m/s and the
temperature by less than 1.7~K. The largest of these is a heating of the first cells, by up to 1.6~K at
\(|V_s|=100\)~m/s. At \(x=65\)~mm the differences are smaller still: below \(y=1.5\)~mm the mean pressure differs
by at most 0.17 percent, the streamwise velocity by at most 0.70~m/s, and the temperature by at most 0.37~K
[Fig.~\ref{fig:avsmeanflow}(d)--(f)].

\section{Derivation of the Speed Distribution of Particles Reflected from the AVS}
\label{sec:derivation}

To find the speed probability distribution of particles reflected from a moving surface, we start with the initial integral:
\begin{equation}
1 = \int_{-\infty}^{\infty} \int_{-\infty}^{\infty} \int_{A\,=\,0}^{\infty} f(V_x) f(V_y) f(V_N) \, dV_x\, dV_y \, dV_N
\end{equation}
where \(f(V_x)\), \(f(V_y)\), and \(f(V_N)\) are the probability distribution functions for the velocities in the x, y, and face normal directions, respectively, and \(A\) is the wall normal surface velocity, which is zero for stationary surfaces. The distribution functions are defined as:
\begin{equation*}
\begin{minipage}{0.45\textwidth}
\begin{equation}
f_t(V_{x,y}) = \sqrt{\frac{m}{2 \pi k_B T}} \exp({-\frac{m}{2 k_B T} V_{x,y}^2})
\end{equation}
\end{minipage}
\hfill
\begin{minipage}{0.45\textwidth}
\begin{equation}
f_n(V_N) = \frac{m}{k_B T} V_N \exp({-\frac{m}{2 k_B T} V_N^2})
\end{equation}
\end{minipage}
\end{equation*}
where \(f_t(V_{x,y})\) describes the tangential velocity components and \(f_n(V_N)\) describes the normal velocity of the molecules leaving the wall, weighted by their flux.
Rewriting the initial integral using these distribution functions, we have:
\begin{equation}
1 = \int_{-\infty}^{\infty} \int_{-\infty}^{\infty} \int_{A\,=\,0}^{\infty} \frac{1}{(2 \pi)} \left( \frac{m}{k_B T} \right)^2 V_N \exp({-\frac{m}{2 k_B T}(V_x^2 + V_y^2)}) \exp({-\frac{m}{2 k_B T} V_N^2}) \, dV_x \, dV_y \, dV_N
\end{equation}
Next, we express the speed probability distribution as:
\begin{equation}
1 = \int_{0}^{\infty} f(C) \, dC \quad \text{where} \quad C = \sqrt{V_x^2 + V_y^2 + V_N^2}
\end{equation}
Transforming coordinates from Cartesian to spherical, we get:
\begin{equation}
dV = dV_x \, dV_y \, dV_N \rightarrow dV = C^2 \cos \theta \, d\theta \, d\psi \, dC
\end{equation}
The volume element in spherical coordinates is thus transformed as shown above. For a stationary wall (\(A=0\)), we rewrite the integral:
\begin{equation}
1 = \int_{-\infty}^{\infty} \int_{-\infty}^{\infty} \int_{A\,=\,0}^{\infty} f(V_x) f(V_y) f(V_N) \, dV \rightarrow \int_{0}^{2\pi} \int_{0}^{\frac{\pi}{2}} \int_{0}^{\infty} f(V_x) f(V_y) f(V_N) C^2 \cos \theta \, dC \, d\theta \, d\psi
\end{equation}
This shows the transformation of the integral limits and the inclusion of the spherical coordinate system.
The normal velocity distribution function is shifted for the reflected particles:
\begin{equation}
f_N(V_N^*) = f_N(V_N - A)
\end{equation}
Where $V_N^*$ is the normal velocities of the particle reflecting from a moving surface. This shift indicates that the velocities of the reflected particles are shifted right by \(A\) in their probability distributions.
The corresponding integral now becomes:
\begin{equation}
1 = \int_{-\infty}^{\infty} \int_{-\infty}^{\infty} \int_{A}^{\infty} f(V_x) f(V_y) f(V_N-A) \, dV_x\, dV_y \, dV_N
\end{equation}
which will be:
\begin{equation}
1 = \int_{-\infty}^{\infty} \int_{-\infty}^{\infty} \int_{A}^{\infty} \frac{1}{2\pi} B^2 (V_N - A) \exp({-\frac{1}{2}B(V_y^2 + V_{x}^2 + (V_N - A)^2)}) \, dV_y \, dV_{x} \, dV_N
\end{equation}
where \( B = \frac{m}{k_B T} \). Now, performing the transformation to the spherical coordinate system, the integral then becomes:
\begin{equation}
1 = \int_{0}^{2\pi} \int_{\max(A,0)}^{\infty} \int_{\theta_{min}}^{\frac{\pi}{2}} \frac{1}{2\pi} B^2 (C \sin \theta - A) \exp({-\frac{B}{2}(C^2 - 2 C A \sin \theta + A^2)}) C^2 \cos \theta \, d\theta \, dC \, d\psi
\end{equation}
where \( \theta_{min} = \arcsin \left( \frac{A}{C} \right) \) for \(C \geq |A|\), and \(\theta_{min} = -\pi/2\) for \(C < |A|\), which occurs only when \(A<0\). For \(A>0\) no reflected molecule is slower than \(A\), so \(f(C)=0\) for \(C<A\) and the speed integral starts at \(A\).
Simplifying the integral for \( B = \frac{m}{k_B T} \) we have:
\begin{equation}
f(C) = C \exp({-\frac{B}{2}(C^2 + A^2)}) \left[ \frac{\exp({ABC}) \left( ABC - A^2 B - 1 \right) + \exp({A^2 B})}{A^2} \right]
\end{equation}
where \( f(C) \, dC \) represents the speed probability distribution of the particles reflecting from a moving wall with positive velocity A in the wall normal direction.
\begin{equation}
1 = \int_{A}^{\infty} f(C) \, dC
\end{equation}
While only the normal velocity component is directly altered by the wall interaction, this change modifies the total particle speed. The particle speed is a direct measure of its kinetic energy, and this energy is then redistributed throughout the flow via subsequent inter-particle collisions, propagating the disturbance introduced by the AVS. The particle speed \( C \) is defined as:
$C = \sqrt{(V_x^*)^2 + (V_y^*)^2 + (V_n^{AVS})^2}$
where \( V_x^* \) and \( V_y^* \) are the tangential velocity components and \( V_n^{AVS} \) is the normal velocity component altered by the presence of the AVS (see Eq.~\eqref{eq:my_equationAVS} of the paper). The resultant speed probability distribution function for particles reflecting from a moving wall takes the following form:
\begin{equation}
    1=\int_{max(V_s^*,0)}^{\infty} f(C) \, dC\quad
\end{equation}
\begin{equation}
\label{eq:maxwellMS}
f(C)=
\begin{cases}
\begin{aligned}
	&\frac{C}{{V_s^*}^2} \exp\!\left(-\frac{B}{2}(C^2+{V_s^*}^2)\right) \\
	&\quad \times \biggl(\exp(B V_s^* C)(B V_s^* C - {V_s^*}^2 B - 1) + \exp({V_s^*}^2 B)\biggr)
\end{aligned} & C \geq |V_s^*| \\[2ex]
\begin{aligned}
	&\frac{C}{{V_s^*}^2} \exp\!\left(-\frac{B}{2}(C^2+{V_s^*}^2)\right) \\
	&\quad \times \biggl(\exp(B V_s^* C)(B V_s^* C - {V_s^*}^2 B - 1) \\
	&\qquad + \exp(-B V_s^* C)(B V_s^* C + {V_s^*}^2 B + 1)\biggr)
\end{aligned} & C < |V_s^*|,\ V_s^* < 0
\end{cases}
\end{equation}
where \( C \) is the particle speed and \( V_s^* \) is the face velocity of the wall and \( B = \frac{m}{k_B T} \), derived above with \(A = V_s^*\). In the limit that \( V_s^* = 0 \), Eq.~\eqref{eq:maxwellMS} becomes:
\begin{equation}
\label{eq:halfmax}
	F(C) = \frac{1}{2} \left( \frac{m}{k_B T_w} \right)^2 C^3 \exp\left( -\frac{mC^2}{2k_B T_w} \right), \quad 0 < C < \infty
\end{equation}
which is the flux-weighted speed distribution of molecules diffusely reflected from a stationary wall.

\begin{figure}[H]
\centering
\includegraphics[width=0.98\linewidth]{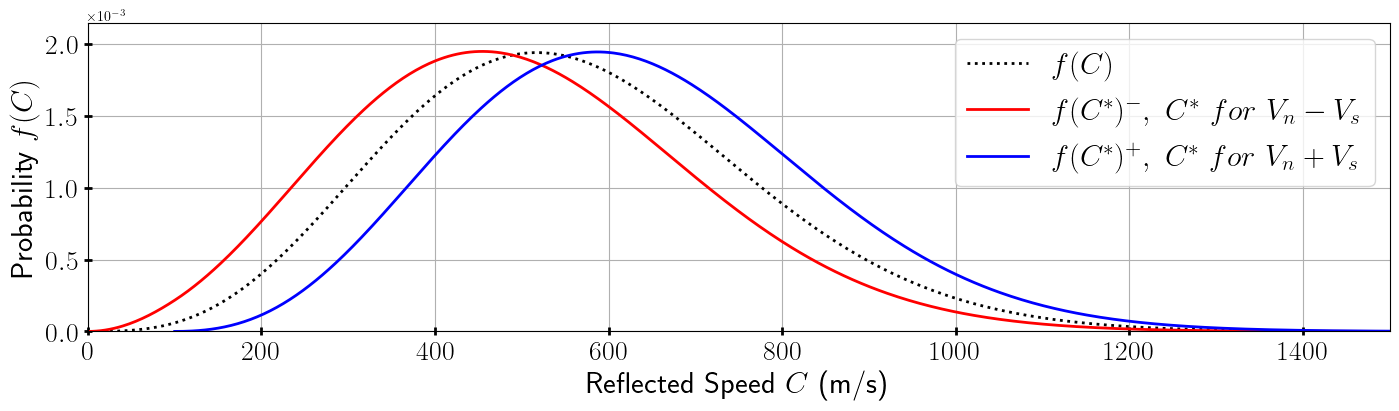}
\caption{Analytic probability distributions of reflected particle speeds, wall temperature of 300~K, with 100~m/s face velocity}
\label{fig:VDF_PM_speed}
\end{figure}

For an AVS with a harmonic face velocity \( |V_s| = 100\, m/s \) and a surface temperature of 300~K, Fig.~\ref{fig:VDF_PM_speed} shows the range of speed distributions for particles reflecting off this vibrating surface. The harmonic face velocity of the AVS introduces clear changes in the overall reflected particle speed distributions. These speed distributions will no longer be steady, and instead, will fluctuate in time due to the AVS's motion. This unsteadiness in particle speeds will propagate throughout the flow via subsequent particle collisions, driving the flow response to the AVS.

\end{document}